\documentclass[10pt,preprint]{emulateapj}

\usepackage{CJK}

\usepackage{amsmath}

\shorttitle{Age discrepancies in Upper Sco}
\shortauthors{Fang et al.}

\begin{document}
\begin{CJK*}{UTF8}{gbsn}

\title{Age spreads and the temperature dependence of age estimates in Upper Sco}
\author{Qiliang Fang(方其亮)\altaffilmark{1,2}, Gregory Herczeg(沈雷歌)\altaffilmark{1},
  Aaron Rizzuto\altaffilmark{3}}
\altaffiltext{1}{Kavli Institute for Astronomy and Astrophysics, Peking University, Yiheyuan Lu 5, Haidian Qu, 100871 Beijing, People's Republic of China}
\altaffiltext{2}{Department of Astronomy, School of Physics, Peking University, 100871 Beijing, China}
\altaffiltext{3}{Department of Astronomy, University of Texas, 2515 Speedway, Stop C1400, Austin, TX 78712, USA}

\begin{abstract}
Past estimates for the age of the Upper Sco Association are typically 11--13 Myr for intermediate-mass stars and 4--5 Myr for low-mass stars.  In this study, we simulate populations of young stars to investigate whether this apparent 
dependence of estimated age on spectral type may be explained by the star formation history of the association.  Solar and intermediate mass stars begin their pre-main sequence evolution on the Hayashi track, with fully convective interiors and cool photospheres.  Intermediate mass stars quickly heat up and transition onto the radiative Henyey track.  As a consequence, for clusters in which star formation occurs on a similar timescale as the transition from a convective to a radiative interior, discrepancies in ages will arise when ages are calculated as a function of temperature instead of mass.  Simple simulations of a cluster with constant star formation over several Myr may explain about half of the difference in inferred ages versus photospheric temperature;  speculative constructions that consist of a constant star formation followed by a large supernova-driven burst could fully explain the differences, including those between F and G stars where evolutionary tracks may be more accurate.  The age spreads of low-mass stars predicted from these prescriptions for star formation are consistent with the observed luminosity spread of Upper Sco.  
The conclusion that a lengthy star formation history will yield a temperature dependence in ages is expected from the basic physics of pre-main sequence evolution and is qualitatively robust to the large uncertainties in pre-main sequence evolutionary models.
\end{abstract}
 

%
%


\section{INTRODUCTION}
Nearby clusters of young stars are our laboratories to study the formation and early evolution of stars and planets.  Masses and ages of young stars in these clusters
are usually measured by isochrone fitting, i.e.~comparing their location on Hertzsprung-Russell (HR) diagram and evolution models. In any given cluster, an accurate stellar locus should provide a robust test for star formation theories.  However, star formation histories and  age spreads are challenging to measure empirically because of observational errors, including unresolved binarity, stellar spots, and membership.  These observational errors are exacerbated by uncertainties in the input physics of pre-main sequence evolution, including birthlines, accretion histories, and magnetic activity \citep[see reviews by][]{soderblom14,hartmann16}, and by inconsistencies between models of stellar evolution.  

One of the nearest young clusters, the Upper Sco Association (Upper Sco), has emerged as a test-bed for pre-main sequence evolution \citep{degeus89,dezeeuw99}.  The temperatures and luminosities of the low-mass\footnote{In this paper, low-mass stars refers to stars from $0.2$ to 1.0 M$_\odot$, with spectral types K to mid-M at the age of Upper Sco.  Intermediate-mass stars refers to AFG stars, with masses $1.0$ to 5 M$_\odot$, while B-type and MS-turnoff stars in Upper Sco are 5-15 M$_\odot$} stellar population in Upper Sco indicate an age of $\sim 5$ Myr \citep[e.g.][]{preibisch02,slesnick08,rizzuto15}.  However, analyses of intermediate and of high mass stars in Upper Sco yield an older age of 10--12 Myr \citep{pecaut12}.  Subsequent studies of low-mass populations confirm these age discrepancies between intermediate- and high-mass stars and low-mass stars \citep{herczeg15,rizzuto16,pecaut16}. Other regions show similar differences, with intermediate-mass stars and fits of empirical isochrones to empirical isochrones typically appearing two times older than K and M-type stars \citep[e.g.][]{hillenbrand97,hillenbrand08,naylor09,malo14,bell15}.

Some of this discrepancy may be resolved with improved calculations of pre-main sequence evolution.  In the most recent models, prescriptions for convection have been updated from the historical assumption of a single, constant mixing length. \citet{feiden16} and \citet{macdonald17} introduce strong magnetic fields in the stellar interiors to slow the efficiency of convection and thereby also slow the stellar contraction rates, yielding inflated radii and therefore older ages for all low-mass pre-main sequence stars \citep[see, e.g.,][]{somers17}.  These models reduce but do not eliminate temperature-dependent discrepancies in age estimates of Upper Sco, and also introduce challenges when applied to other regions (see Appendix A).  Similar reductions in convective efficiency have led to older ages when photospheric spots are considered \citep{somers15} or when the treatment of convection is based on results from 3D MHD simulations \citep{baraffe15}.  The uncertainty in ages related to accretion history would preferentially make low-mass stars less luminous \citep[e.g.][]{hartmann97,baraffe16}, thereby exacerbating the age discrepancy.  Uncertainties in stellar atmospheres caused by errors in opacities \citep[e.g.][]{bell13} and boundary conditions \citep[e.g.][]{chen14} also propagate into errors in the evolutionary tracks and in the placement of stars in HR diagrams.

We investigate an alternative explanation, that the conflict in ages between cool and hot stars may be explained with basic stellar evolution and an age spread within the cluster.   
During their pre-main sequence evolution, low-mass young stars are fully convective and evolve along the Hayashi track, with a roughly constant temperature and a luminosity that decreases as the star contracts.  On the other hand, intermediate-mass stars begin their lives fully convective, but the stellar interior heats up, thereby reducing the opacity and allowing for efficient radiative energy transport \citep{hayashi61}.  Their evolution then proceeds along the roughly-horizontal Henyey track, during which the stellar photosphere becomes hotter while maintaining a similar luminosity \citep{henyey55,hayashi65}.  If the star formation timescale in a cluster is similar to the timescale for intermediate mass stars to become radiative (1-10 Myr), then measuring ages as a function of photospheric temperature \citep[e.g.][]{pecaut12,herczeg15} will mean that hotter (spectral types AFG) stars will tend to be the older stars evolving on the Henyey track, while younger stars with the same mass will still be on the Hayashi track and will have cool effective temperatures (K and early M spectral types).  For young ($\lesssim 30$ Myr) clusters, the AFG stars will preferentially be among the oldest stars in the cluster, while M stars will sample the full star formation history.

These effects are likely broadly applicable to young regions. Some age spread is required by the presence of embedded objects, objects with disks, and diskless objects within the same cluster \citep[e.g.][]{evans09}.  A halo of mostly-diskless young stars spread about between dense molecular clouds in the Taurus star forming region suggests either a large age spread or past bursts of star formation \citep{kraus17}.  While concentrated regions may form stars quickly, the measured age spread should be expected to be longer when the region considered is larger (see, e.g., description of Chamaeleon by \cite{sacco17} and age spreads of tens to hundreds of Myr in massive clusters inferred by, e.g., \cite{goudfrooij11}).

In this paper, we model young clusters with a range of star formation histories and then apply the results to Upper Sco, which has evidence for a spatial age gradient \citep{preibisch07,pecaut16} and an age spread from HR diagrams, including B-type MS-turnoff stars with ages that range from 5--14 Myr  \citep[e.g.][]{pecaut12,rizzuto15}.  In \S 2, we first establish the census of probable Upper Sco members from the literature to create a stellar locus.   In \S 3, we use stellar evolution models to simulate HR diagrams to establish how age depends on temperature for a cluster with an age spread.  In \S 4, we then use apply these simulations to Upper Sco to evaluate whether star formation history may lead to a temperature dependence in age estimates.   We discuss these results in \S 5.






\section{THE POPULATION OF UPPER SCO}

In this section, we compile a list of high-probability members of Upper Sco based on past selection criteria, proper motion, photometry, and distance, and then characterize their luminosities and photospheric temperatures.  Our selection criteria excludes members with anomalous proper motions or distances to minimize interlopers.  We also exclude stars with disks.  These selection criteria should lead to an accurate measurement of the locus of Upper Sco stars in the HR diagram.  

The methods described for selecting members and assigning temperatures and luminosities are similar to, and in some cases follow exactly, the methods implemented for Upper Sco by \citet{pecaut16}.  Our sample, adopted from several literature studies, extends to M stars, while \citet{pecaut16} concentrates on K stars.  We also adopt a single distance for stars without parallax measurements, while \citet{pecaut16} use distance estimates from kinematic analyses.  Relative to \citet{pecaut16}, our sample of stars cooler than 4000 K is more complete and less luminous.

\subsection{Sample Description}

The initial membership list of Upper Scorpius and spectral type measurements is compiled from \citet{rizzuto15}, \citet{pecaut16} and the compilation of \citet{luhman12}, which included previous surveys of \citet{preibisch99}, \citet{ardila00}, \citet{preibisch02}, and \citet{slesnick06usco}.  Photometric data for low mass members of Upper Sco are obtained from the 2MASS $JHKs$ survey \citep{skrutskie06}, WISE $W1$, $W2$, $W3$, $W4$ survey \citep{cutri13}, and APASS \citep{henden12}.  Proper motions in this work are compiled from GAIA DR1 \citep{gaia2016dr}, UCAC4 \citep{zacharias13}, PPMXL \citep{roeser10}, and SPM4 surveys \citep{girard11}.   We focus on members with spectral types that range from F0--M5.

\begin{figure}[!t]
\includegraphics[width = 8cm]{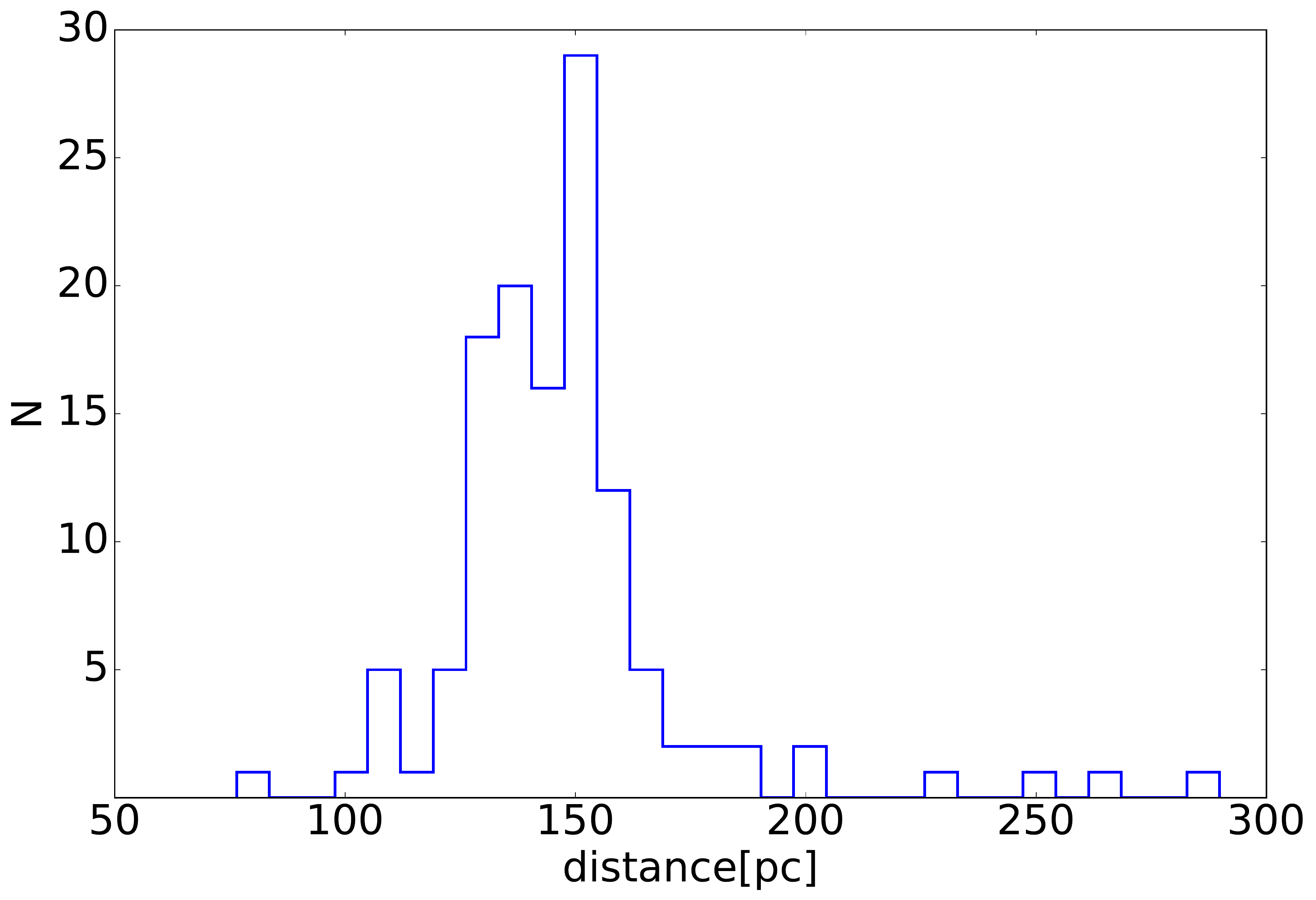}
\centering
\caption{The distribution of distances of members in Upper Scorpius compiled from GAIA catalog, with a mean of 144.2 pc and a standard deviation of 17.6 pc.}
\label{fig:distances}
\end{figure}

The \citet{gaia2016dr} DR1 TGAS astrometric catalog includes parallaxes of 126 F, G, and K-type objects that were previously identified as members of Upper Sco and are located between $234$ to $253$ degrees in right ascension and $-34$ to $-14$ degrees in declination.  Figure ~\ref{fig:distances} shows that the distance distribution of these members has a mean of 144 pc, with a standard deviation of 17.6 pc, a mean error of 8.6 pc, and a systematic error of $\sim 6$ pc \citep{arenou17}.  A quadratic subtraction of these two errors indicates that the distance spread is $\sim 15$ pc, consistent with previous estimates \citep[e.g.][]{slesnick08}.  Three stars with distances smaller than 100 pc (10 mas in parallax) or larger than 200 pc (5 mas in parallax) are rejected as members (Table~\ref{tab:rejections.tab}).  We adopt distances from the Gaia DR1 parallaxes when available, or the average distance of $145\pm15$ pc for stars without parallax measurements.

Some members of the older Upper Centaurus-Lupus subgroup, a related association also within the Sco-Cen OB Association, likely contaminate this membership list.  The related moving groups have similar space motions and overlap on the sky \citep{dezeeuw99,pecaut12}.  Some contamination is also expected from the younger $\rho$ Ophiucus star forming region.  

\begin{table}[!b]
\caption{Rejected members included in \citet{pecaut12}}
\label{tab:rejections.tab}
\begin{tabular}{lccc}
\hline
2MASS&SpT&$\pi_{GAIA} [mas]$&\\
 \hline
J15545986-2347181&G3V&13.08 $\pm$ 0.68\\ 
J15582930-2124039&F0IV&3.45 $\pm$ 0.26\\ 
J16123604-2723031&K4&1.03 $\pm$ 0.88\\ 
 \hline%
 \end{tabular}
\end{table}

\begin{figure}[!t]
\includegraphics[width = 8cm]{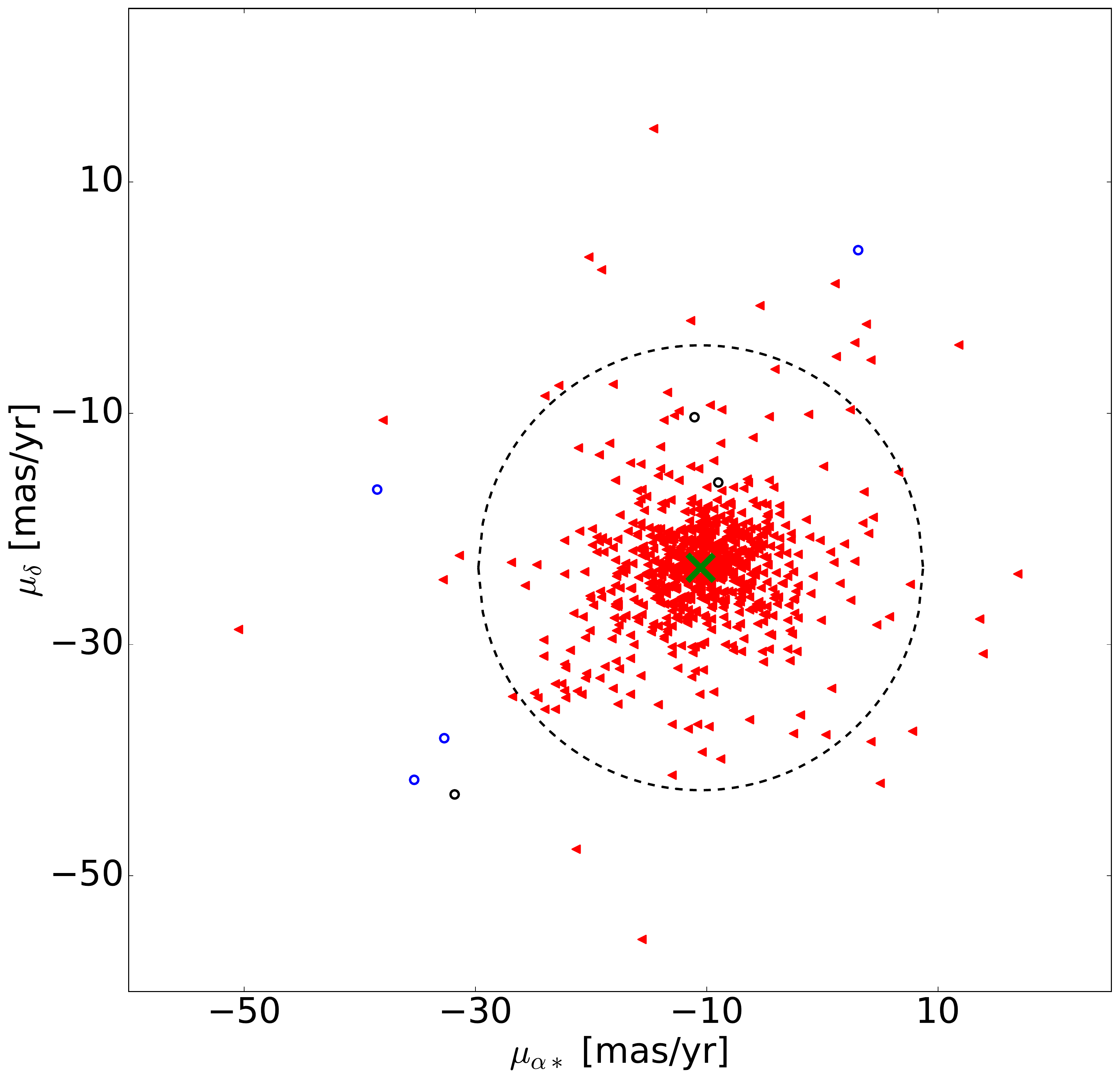}
\centering
\caption{The proper motion of candidate members of Upper Scorpius. The red triangles are accepted members, the blue dots are objects rejected by proper motion selection criteria, the black dots are objects rejected by parallax and the green cross indicates the mean value of proper motions. The black dashed circle indicates the 3$\sigma$ spread of proper motions.}
\label{fig:pm1}
\end{figure}

Figure \ref{fig:pm1} shows the proper motions of the candidates in both directions, where $\mu_{\alpha*} \equiv \mu_{\alpha} cos\delta$.  The average proper motion of all Upper Scorpius members considered in this paper is $\mu_{\alpha*} = -10.45$ mas yr$^{-1}$ and $\mu_{\delta} = -23.21$ mas yr$^{-1}$, with a total standard deviation of 6.41 mas yr$^{-1}$. To reject objects that are moving inconsistently with the common streaming motion, all stars with proper motions that deviate 3$\sigma$ from the mean are rejected as members. The remaining 767 members are kinematic selected members and will be used in the following analysis.

\subsection{Stellar properties}

In this section, we describe how we combine the literature spectral type information with existing photometry to measure stellar properties and identify stars with disks, which are subsequently excluded from analysis.

\subsubsection{Extinction}
We adopt the conversion between spectral type, temperature, and color scales from \citet{pecaut13}.  Uncertainties in temperatures are assigned to be 200 K at temperatures $>4100$ K, 125 K at  3750 -- 4100 K, and 50 K below 3750 K.   Following \citet{pecaut16}, for stars with $V$-band photometry, extinction is estimated from the color excesses $E(V - J)$, $E(V - H)$, $E(V - K)$ relative to the expected photospheric color. for other sources the extinction is estimated from the $E(J-K)$ color excess.    The color excesses are then converted to extinction using  $A_J/A_V$ = 0.27, $A_H/A_V$ = 0.17, $A_K/A_V$ = 0.11 \citep{fiorucci03}.  The adopted V band extinction is the mean value of $A_V$ estimated from these colors.

\begin{figure}[!t]
\includegraphics[width = 8cm, height = 13.33cm]{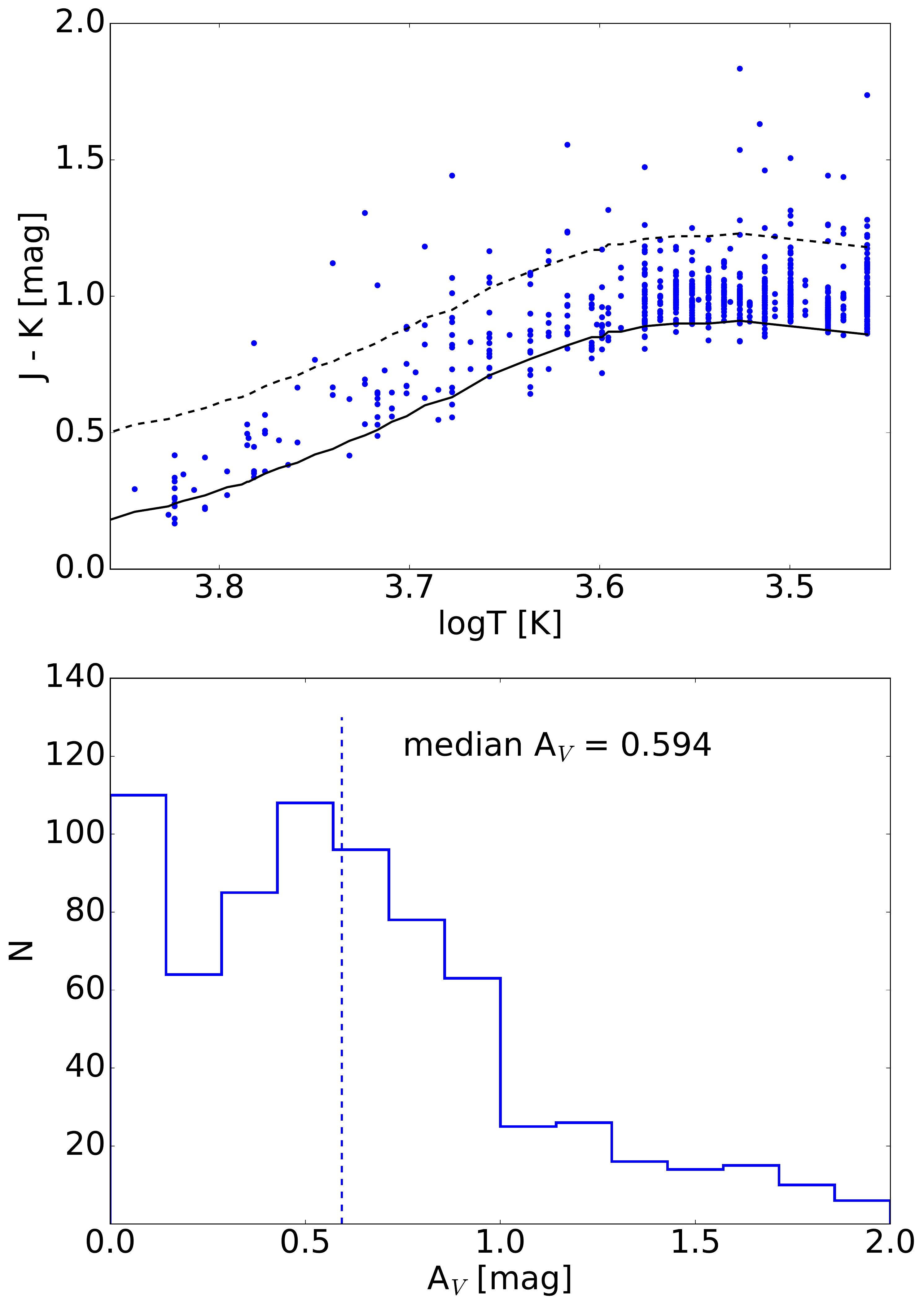}
\centering
\caption{Top panel: Temperatures versus J - K color for diskless members of Upper Sco, compared to the photospheric color (solid line) and the photospheric color that corresponds to $A_V=2$ mag (dashed line).   Bottom panel:  The distribution of estimated extinction.  Negative $A_V$ are unphysical and assumed to be 0 mag, while stars with $A_V > 2$ mag are rejected.}
\label{extinction}
\end{figure}

The top panel of Figure~\ref{extinction} shows the $J - K$ color for low mass members of Upper Sco. The black solid line is the relation between effective temperature and intrinsic color, which is derived from \citep{pecaut13} and the dashed black line is the color reddened by $A_V$ = 2 mag. The median extinction is $A_V \sim 0.60$ mag. All objects reddened by $A_{V} > 2$ mag are excluded from further analysis.  All extinctions $A_V < 0$ mag are set to $A_V = 0$ mag.   Use of negative extinctions would have a negligible affect on the empirical isochrone of Upper Sco.  The median $A_V$ is constant versus spectral type to within $0.25$ mag, which leads to systematic errors of $<7$\% in luminosity.

\subsubsection{Disk identification}
\begin{figure}[!t]
\includegraphics[width = 8cm, height = 6.22cm]{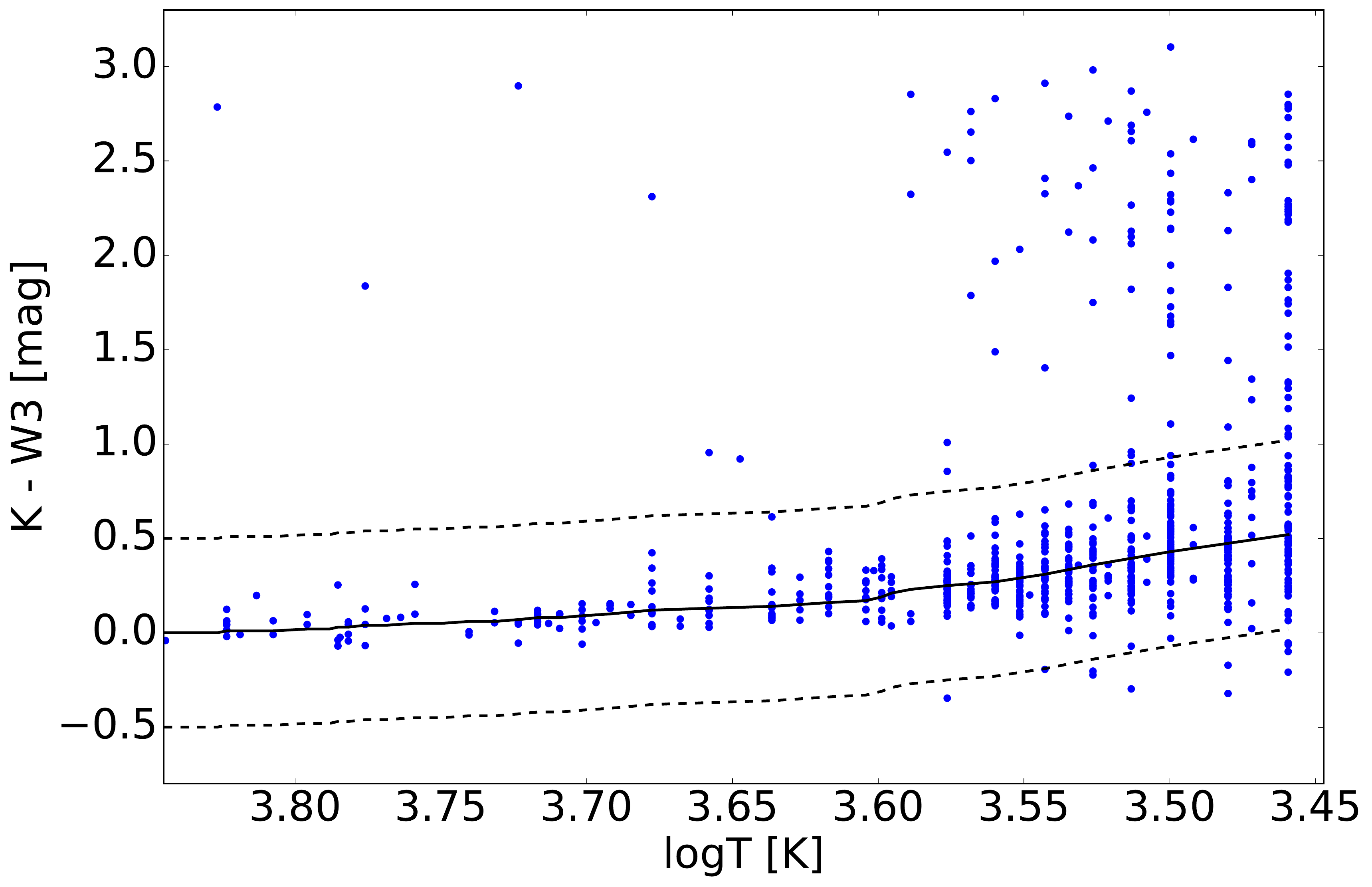}
\centering
\caption{Temperature versus $K-W3$ color.  The  color and the dashed line is 0.5 mag offset. Stars with $K - W3$ more than 0.5 mag above the median value (solid line is median, dashed lines show $\pm0.5$ from the median) likely have disks and are excluded from our analysis.  Stars with $K-W3$ of 0.5 mag below the median value are also excluded for consistency.}
\label{fig:Disk}
\end{figure}

Disks produce excess emission in the infrared from dust and in the optical from accretion onto the star. Since correcting photometry for these processes is challenging, we instead identify and exclude all disk sources in our sample.  In this work, disks are identified based on a mid-IR color excess, with $E(K-W3)>0.5$ \citep[see also, e.g.,][]{luhman12}, as shown in Figure ~\ref{fig:Disk}.  These results are consistent with the disk identification in Upper Sco by \citet{luhman12} and \citet{rizzuto15}.  Uncertainty in the $K-W3$ color leads to a higher rejection rate of diskless stars at $<3000$ K.

\subsubsection{Luminosity}
Luminosities are calculated from 2MASS $J$-band  photometry and the bolometric correction of \citet{pecaut13} obtained for the appropriate spectral type, as follows:
\begin{equation}
  \log \frac{L}{L_{\odot}} = -0.4 (m_J - 5\log\frac{d}{10 pc} + BC_J - M_{bol, \odot})
\label{eq:luminosity}
\end{equation}
where the bolometric luminosity of the Sun, $M_{bol, \odot}$, is 4.755 mag (Mamajek 2012).
The error of luminosity, $\sigma L$, is estimated as
\begin{equation}
  \sigma(\log \frac{L}{L_{\odot}}) = 0.4 \sqrt[]{\sigma (m_J)^2 + [\frac{5\sigma(d)}{d \rm ln 10}]^2 + \sigma (BC_J)^2}
 \label{eq:error_luminosity}
\end{equation}
Estimating the first two terms is straightforward, as $\sigma$(m$_J$) is the observation error of the $J$ band magnitude and $\sigma$(d) is distance spread (15 pc, see \S 2). To estimate the errors of bolometric corrections, we perform 1000 Monte Carlo simulations for each star. In each simulation, we assign the same error in temperature as discussed in previous sections, and derived a set of bolometric corrections for each star. The standard deviation is the adopted error in the third term in Equation~\ref{eq:error_luminosity}.  The error in bolometric magnitude then propagates to the error in luminosity.

\subsection{HR diagram of Upper Sco}

\begin{figure}[!t]
\includegraphics[width = 8cm, height = 6.22cm]{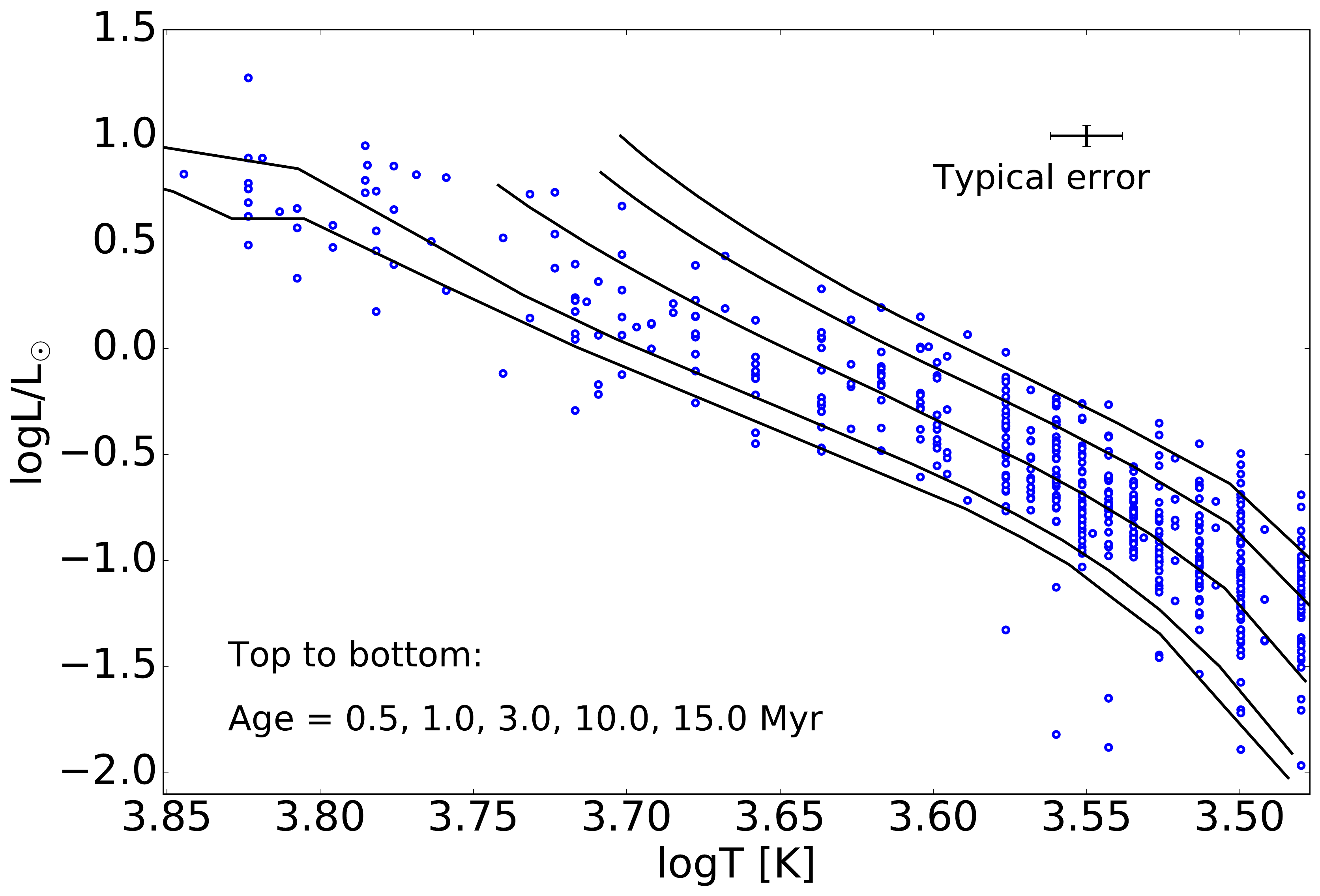}
\centering
\caption{The HR diagram of diskless pre-main sequence stars in Upper Sco. Black solid lines are \citet{dotter08} theoretical isochrones at 1, 3, 5, 10, and 15 Myr from top to bottom.}
\label{fig:hrdiag}
\end{figure}

Figure~\ref{fig:hrdiag} shows the temperatures and luminosities of the diskless members of Upper Sco in an HR diagram.  
The empirical isochrone is calculated by dividing the sample into $\log T_{\rm eff}$ ranges from 3.5 (3000 K) to 3.9 (7000 K) into 20 bins of $\sim 0.02$ dex.  Bins with fewer than three stars are ignored.  All stars that are located either above the 0.1 Myr isochrone or below the main sequence are rejected as non-members. 
Errors of temperatures and luminosities are assigned as described above. For each star, the adopted value and uncertainty for age is evaluated from ages obtained in $10^4$ Monte Carlo simulations in which these 1 $\sigma$ uncertainties are randomly applied to the effective temperature and luminosity. The empirical isochrone is then defined as the median of all luminosities within the bin.  
As has been established in other studies \citep[e.g.][]{pecaut12,rizzuto15,herczeg15,pecaut16}, the average age of hotter stars is older than that of younger stars.

\begin{figure*}[!t]
\includegraphics[width = 18cm, height = 5.5cm]{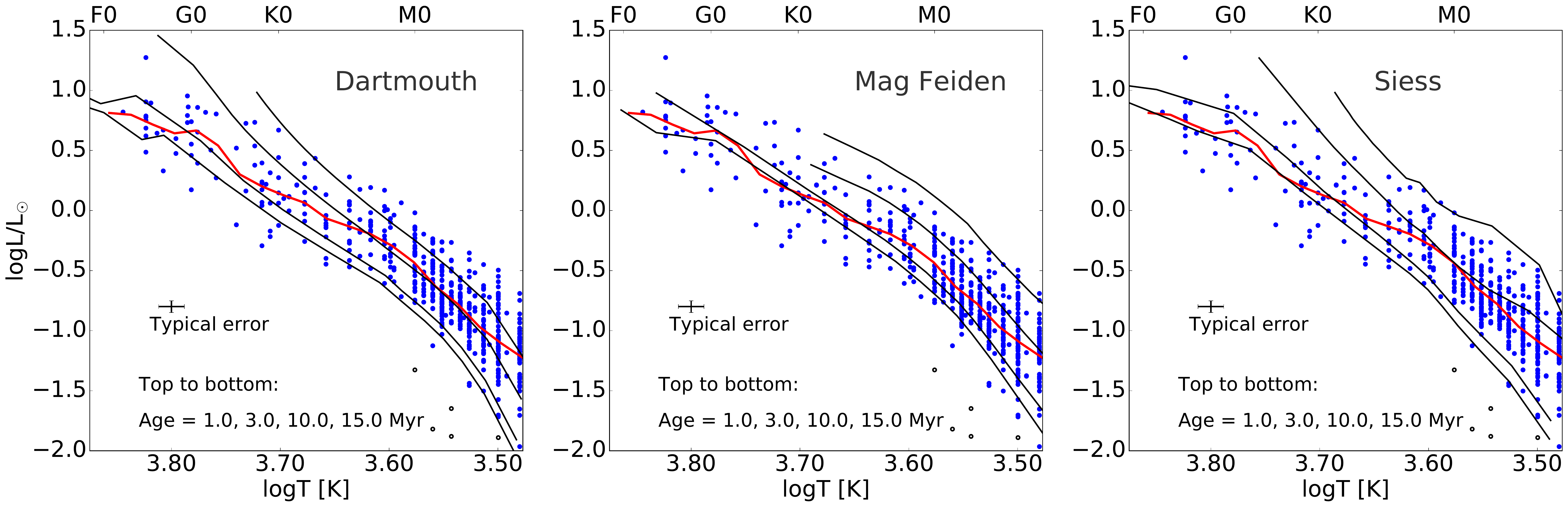}
\centering
\caption{Emperical isochrone of Upper Sco members compared with the model isochrones from the Dotter, the magnetic Feiden, and Siess pre-main sequence evolutionary models, with isochrones of 1, 3, 10, and 15 Myr.  Compared to the isochrone of \citet{pecaut16} in their similar Figure 5, the isochrone here is similar for stars hotter than 4000 K and is fainter for cooler stars.}
\label{fig:isochrone1}
\end{figure*}

Figure~\ref{fig:isochrone1} compares the empirical isochrone with stellar evolution tracks\footnote{An overview of many pre-main sequence evolutionary models can be found in \citet{stassun14}.  The Siess and Dotter tracks use standard mixing length theory, while Feiden introduces magnetic fields into the stellar interior.} from \citet[][hereafter Siess]{siess00}, \citet[][hereafter Dotter]{dotter08} and the magnetic tracks of \citet[][hereafter Feiden]{feiden16}. 
The isochrone of Upper Scorpius is not parallel to any of the theoretical isochrones, as should be expected for co-eval members of a stellar cluster (if pre-main sequence evolutionary models are accurate).
The age measurement strongly depends on temperature in all models for pre-main sequence evolution (see median age of each spectral bin in Table \ref{tab:ages.tab}). The empirical isochrone calculated here for Upper Sco is very similar to that of \citet{pecaut16} at temperatures hotter than 4000 K and is several tenths of dex fainter at cooler temperatures.

This sample suffers from several biases.  First, the low-mass stellar population is incomplete, with significant spatial biases.  The disk fraction versus luminosity within a small range of subclasses indicates that the sample may be preferentially missing faint targets at spectral types later than M3.
Early-type members were identified by \citet{dezeeuw99} using Hipparcos cover a large area on the sky, while many searches for low-mass stars focus on smaller regions.  The sample also suffers from contamination of stars that belong to nearby, related regions.  Finally, our exclusion of disk sources may preferentially exclude some of the youngest sources in the region. None of these biases affects our comparisons of inferred ages of low-mass and intermediate-mass stars.

\begin{table}[!b]
\begin{center}
\caption{Median age versus spectral type for Upper Sco}
\label{tab:ages.tab}
\begin{tabular}{lccc}
\hline
&&magnetic \\
SpT& Dotter & Feiden & Siess\\ \hline
F2& 13.3 $\pm$ 0.7 &15.8 $\pm$ 0.9 &13.9 $\pm$ 1.5\\
F5& 13.6 $\pm$ 1.5 &16.3 $\pm$ 1.5 &14.9 $\pm$ 1.4\\ 
F8& 10.8 $\pm$ 2.0 &13.3 $\pm$ 3.3 &14.0 $\pm$ 1.7\\ 
G2& 8.6 $\pm$ 2.3 &11.4 $\pm$ 4.7 &13.1 $\pm$ 2.4\\ 
G5& 8.7 $\pm$ 3.0 &16.0 $\pm$ 6.6 &14.8 $\pm$ 4.0\\ 
G8& 7.9 $\pm$ 1.5 &16.1 $\pm$ 3.3 &13.5 $\pm$ 2.1\\ 
K2& 5.5 $\pm$ 1.2 &11.4 $\pm$ 2.7 &9.5 $\pm$ 1.6\\ 
K5& 2.7 $\pm$ 0.5 &7.3 $\pm$ 1.1 &4.1 $\pm$ 0.7\\ 
K8& 2.2 $\pm$ 0.3 &6.4 $\pm$ 0.7 &3.4 $\pm$ 0.4\\
M0& 2.2 $\pm$ 0.4 &5.9 $\pm$ 0.8 &3.0 $\pm$ 0.4\\ 
M2& 2.9 $\pm$ 0.2 &6.2 $\pm$ 0.4 &3.7 $\pm$ 0.2\\
\hline
\end{tabular}
\end{center}
\end{table}

\section{POPULATION SYNTHESIS OF YOUNG STAR CLUSTERS}

If a population of young stars forms over some duration of time, then stars with hotter photospheres will be preferentially older than stars with cooler photospheres. In the Dotter evolutionary models, at 5 Myr 
the mass range of F type (here defined as $6000-7500$ K) stars is 1.95 $\sim$ 2.14 M$_{\odot}$, while at 10 Myr same temperature range corresponds to masses of  1.50 $\sim$ 1.79 M$_{\odot}$. By integrating over the Chabrier initial mass function \citep{chabrier03}, a 10 Myr old population will have five times more F-type stars than a 5 Myr old population.  On the other hand, the 10 Myr old population will have 1.5 times fewer K-type stars (here defined as 4000-5000 K) than a 5 Myr old population.  In this context, a star cluster that forms over 10 Myr will have a measured age that depends on the spectral type or temperature range being measured.  The median age estimated for F type stars will be biased to older stars, while the age estimated for K type stars will be biased to younger stars.

In this section, we quantify the size of this effect and characterize the age spread and star formation history that would be needed to reproduce the ages of the population in Upper Sco.  We first simulate HR diagrams in a cluster to derive model isochrones to compare with the observed isochrone of Upper Sco under some simple assumptions.  We then introduce some speculative changes that are able to entirely reproduce the empirical isochrones and age spread of Upper Sco.

\subsection{Simulation setup}

Stellar populations are simulated using the pre-main sequence evolutionary models of \citet{dotter08}.
A sample of $10^4$ stars with masses $0.1$ to 2.0 M$_\odot$ are generated with masses randomly selected from the Chabrier initial mass function, which describes Galactic disk stars \citep{chabrier03}. We then introduce to each star (or binary system) an age randomly generated from the star formation history.

The main sources of uncertainty are errors in spectral types, photometry, and distance, and in unresolved multiplicity.  In the previous section, we assign to each member in Upper Sco errors in both effective temperature and luminosity. The mean errors in $\log T$/K and $\log L/L_\odot$ are assumed to be in a Gaussian distribution with a $1\sigma$ scatter of 0.01 and 0.05 dex, respectively \citep[see also a similar implementation in][]{jose17}.

Unresolved binaries will lead to incorrect effective temperatures and luminosities, resulting in inferred ages that are too young \citep[e.g.][]{preibisch99,slesnick08,reggiani11}. To estimate the effect of binarity (higher order systems are ignored), the effective temperature of multiple systems are assumed to be that of the primary star, although the combined temperature will be lower than the temperature of the primary alone \citep[e.g., see discussion in][]{herczeg14}. The luminosity of binary systems are estimated by summing the primary and companion luminosity, which will lead to small differences with our observed luminosity measurements from the $J$-band photometry.

We follow the quantification of the multiplicity fraction and mass ratio of \citet{lu13}, which was estimated from the empirical results of \citet{preibisch99,kouwenhoven07,kobulnicky07,lafreniere08}.
The multiplicity fraction adopted here follows power law with the form $MF(m)  =0.44 m^{0.51}$ given by \citet{lu13} for stars with masses less than $5$ M$_\odot$, and where m is the primary mass.  All systems in our work have either one or two stars.
The mass ratio $q\equiv$ $\frac{M_2}{M_1}$ between the companion star and primary star also follow a power law,
\begin{equation}
  p(q) =
    (\frac{1 + \beta}{1 - q{_{lo}}{^{1 + \beta}}})q^\beta
\end{equation}
where $\beta$ ranges from -0.6 to 1.4. In this work we adopt a flat companion mass ratio distribution ($\beta$ = 0), following \citet{reggiani13}.  

The effect of pairing function is carefully discussed in previous work \citep[see, e.g.][]{kouwenhoven09}.  Here we adopt the primary-constrained random pairing method, where the primary mass is derived from Chabrier IMF and the companion mass is determined by $p(q)$.  All companion stars with masses lower than 0.1 M$_\odot$ will not contribute to significant changes in the HR diagram and are excluded from further analysis.
 
\subsection{The Star Formation History of a Synthetic Stellar Population}

In this subsection, we describe how the star formation history is implemented into our population synthesis.  We first calculate stellar populations for the Smooth-Quench model, which has a constant star formation over some time period.  We then calculate populations in the Smooth-Burst model, which adds a burst of star formation at the end of the Smooth-Quench model.  The Smooth-Quench model is the simplest possible characterization of a star formation history.  While {\it ad hoc} additions to the Smooth-Quench model could become hopelessly complicated, adding a burst at the end of the Smooth-Quench model maximizes possible age differences between stars of different temperatures, thereby probing the limits of age differences versus spectral type.
The Smooth-Burst model also may represent a reasonable scenario of a past supernova triggering a burst in star formation (see \S 4).   The accelerating star formation model of \citet{palla00} is not investigated in this work but would yield results that are intermediate between the Smooth-Quench and Smooth-Burst models.

\begin{figure}[!t]
\includegraphics[width = 8cm, height = 6.22cm]{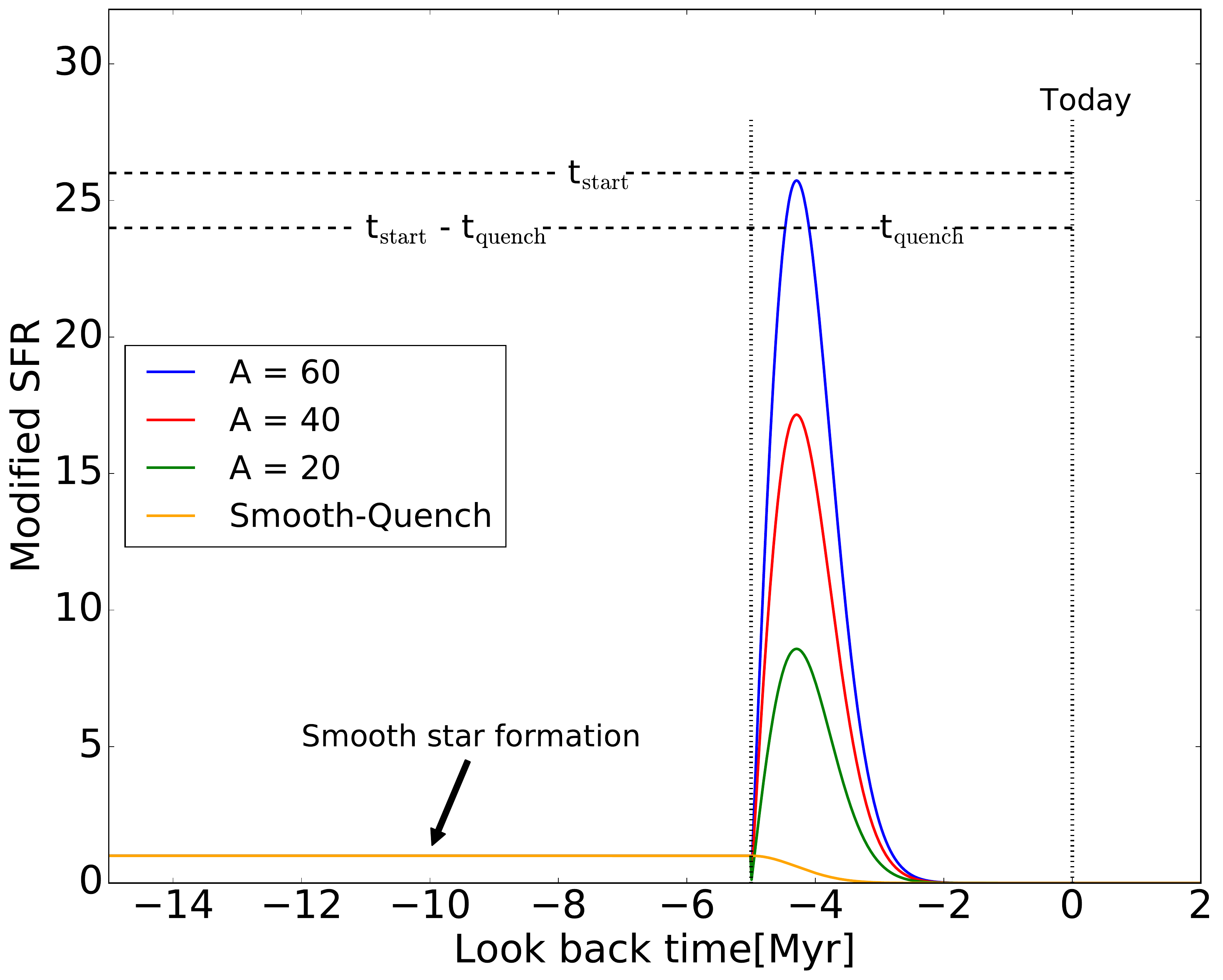}
\centering
\caption{An illustration of the star formation histories adopted in this work.  In this example, the age of the start of star formation in the association, $t_{\rm start}$, is 15 Myr and the time since the start of the burst/quench ($t_{\rm quench}$) is 5 Myr.}
\label{fig:sfhistory}
\end{figure}

These models are summarized in Figure~\ref{fig:sfhistory}.  The Smooth-Quench model is described by two parameters: the time since the association first started forming stars, $t_{\rm start}$, and the time since the star formation was quenched, $t_{\rm quench}$. The decrease in star formation after quenching is described by e$^{-t^2}$. The star formation rate is then analytically described by
\begin{equation}
  SFR(t) =%
  \begin{cases}
    1 &\ t \leq t_{\rm start} - t_{\rm quench} \\
    e^{-[t - (t_{\rm start} - t_{\rm quench})]^2} &\ t_{\rm start} - t_{\rm quench} < t
  \end{cases}
\end{equation}
In addition to $t_{\rm start}$ and $t_{\rm quench}$, the Smooth-Burst model adds a burst of strength $A$ relative to the quiescent star formation rate at time $t_{quench}$. The star formation rate is then described as
\begin{equation}
  SFR(t) =%
  \begin{cases}
    1~~~~~(t \leq t_{\rm start} - t_{\rm quench}) \\
    A[t - (t_{\rm start} - t_{\rm quench})]e^{-[t - (t_{\rm start} - t_{\rm quench})]^2} \\~~~~~~(t_{\rm start} - t_{\rm quench} < t) & \\
  \end{cases}
\end{equation}

For the Smooth-Quench model, by integrating over the star formation history, the ratio of the number of stars form at the quench/burst process N$_1$ and the number of stars form during the constant star formation rate $N_0$ is
\begin{equation}
  \frac{N_1}{N_0} = \frac{0.886}{t_{\rm start} - t_{\rm quench}}
 \label{eq:ratio_quench}
\end{equation}
 and for the Smooth-Burst model
 \begin{equation}
  \frac{N_1}{N_0} = \frac{0.5A}{t_{\rm start} - t_{\rm quench}}
\label{eq:ratio_burst}
\end{equation}
where t$_{start}$ and t$_{quench}$ are both in the unit of Myr.  Thus, if star formation occurs over 10 Myr ($t_{\rm start}-t_{\rm quench}=10$), then an amplitude of the burst $A=20$ would yield an equal number of stars formed during the burst and during the smooth period of star formation.

In both of these constructions, the IMF, binary formation, and stellar evolution are assumed to be constant and independent of star formation rate.  In our models, these parameters are calculated by generating $10^4$ stars over $10^4$ possible ages during the star formation history and randomly assigning these stars a mass from the Chabrier IMF and binarity as described above.

\vspace{10mm}

\subsection{Simulation Results} 
Our simulations are run with several sets of parameters, listed in Table ~\ref{tab:paramter}.
For each simulated population, the median ages of stars are estimated in different temperature ranges, with an empirical isochrone calculated following the same steps in \S 2.  
The empirical isochrones of the simulated HR diagrams from these models are then simplified into two measureable parameters: the age difference between F5 and M0 stars ($\sim 6500$ K and $3900$ K, respectively) for the entire cluster and the luminosity spread of M0 stars. Binaries have only a minor effect in these simulations.

\begin{table}[!b]
\caption{Parameters}
\label{tab:paramter}
\begin{tabular}{lcc}
\hline
&Smooth-Burst & Smooth-Quench \\ \hline
$t_{\rm start}$ & 12, 14, 16, 18, 20 &12, 14, 16, 18, 20\\
$t_{\rm quench}$ (Myr)& 2, 4, 6, 8, 10& 2, 4, 6, 8, 10\\
$A$& 20, 40, 60, 80& -- \\
\hline
\end{tabular}
\end{table}

\begin{figure}[!t]
\includegraphics[width = 8cm, height = 13.33cm]{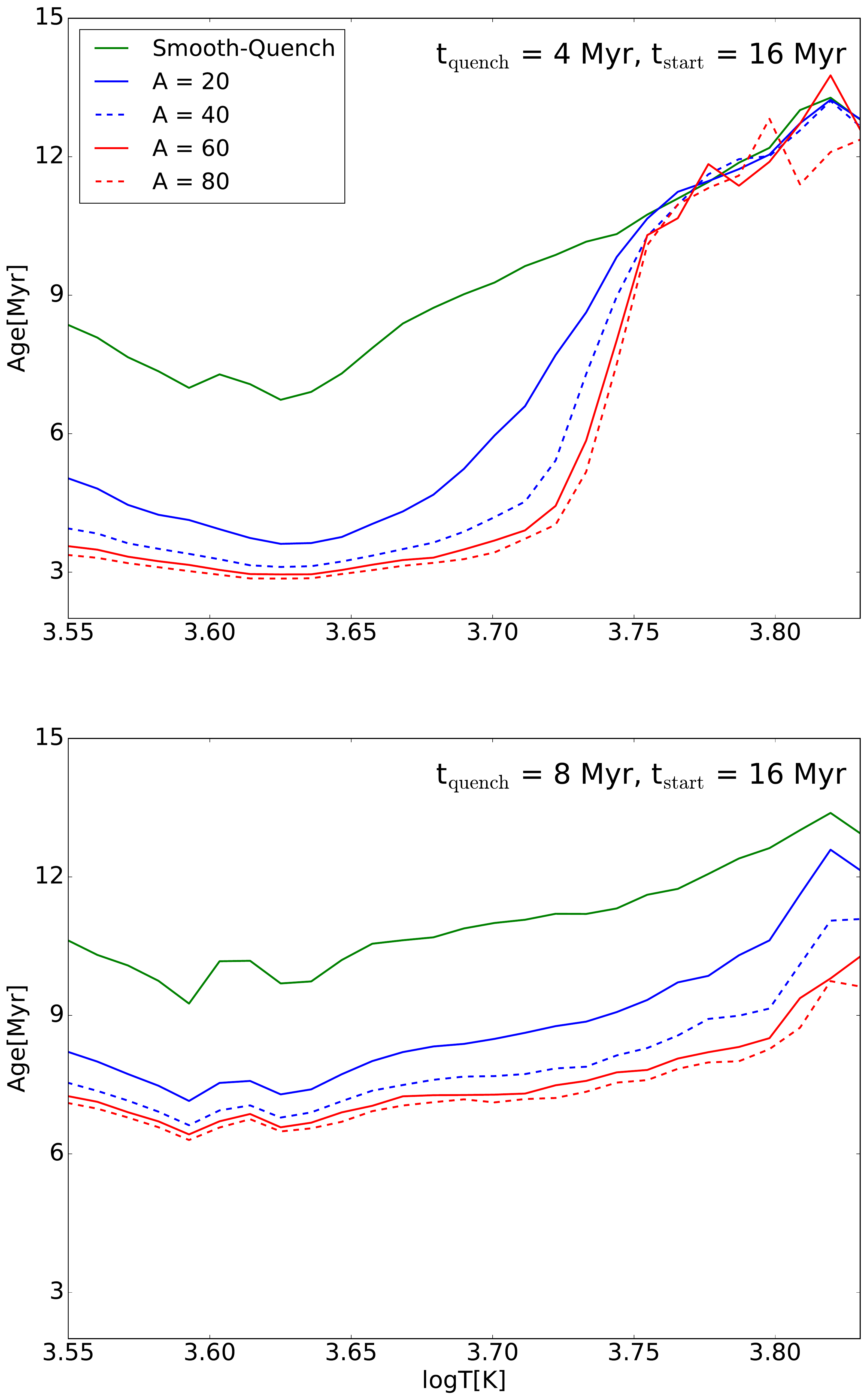}
\centering
\caption{Age versus temperature for the median star in a temperature bin in our simulations.  The top panel shows that age differences are large for a model with $t_{\rm start} = 16$ Myr and $t_{\rm quench} = 4$ Myr, while these differences are reduced for a model with  $t_{\rm start} = 16$ Myr and $t_{\rm quench} = 8$ Myr, i.e., less of an age spread and star formation quenching further in the past. In each panel, simulations with different $A$ values are labeled by different colors and line styles.}
\label{fig:Burst_A_detail}
\end{figure}

\begin{figure*}[!t]
\includegraphics[width = 13.5cm, height = 10.5cm]{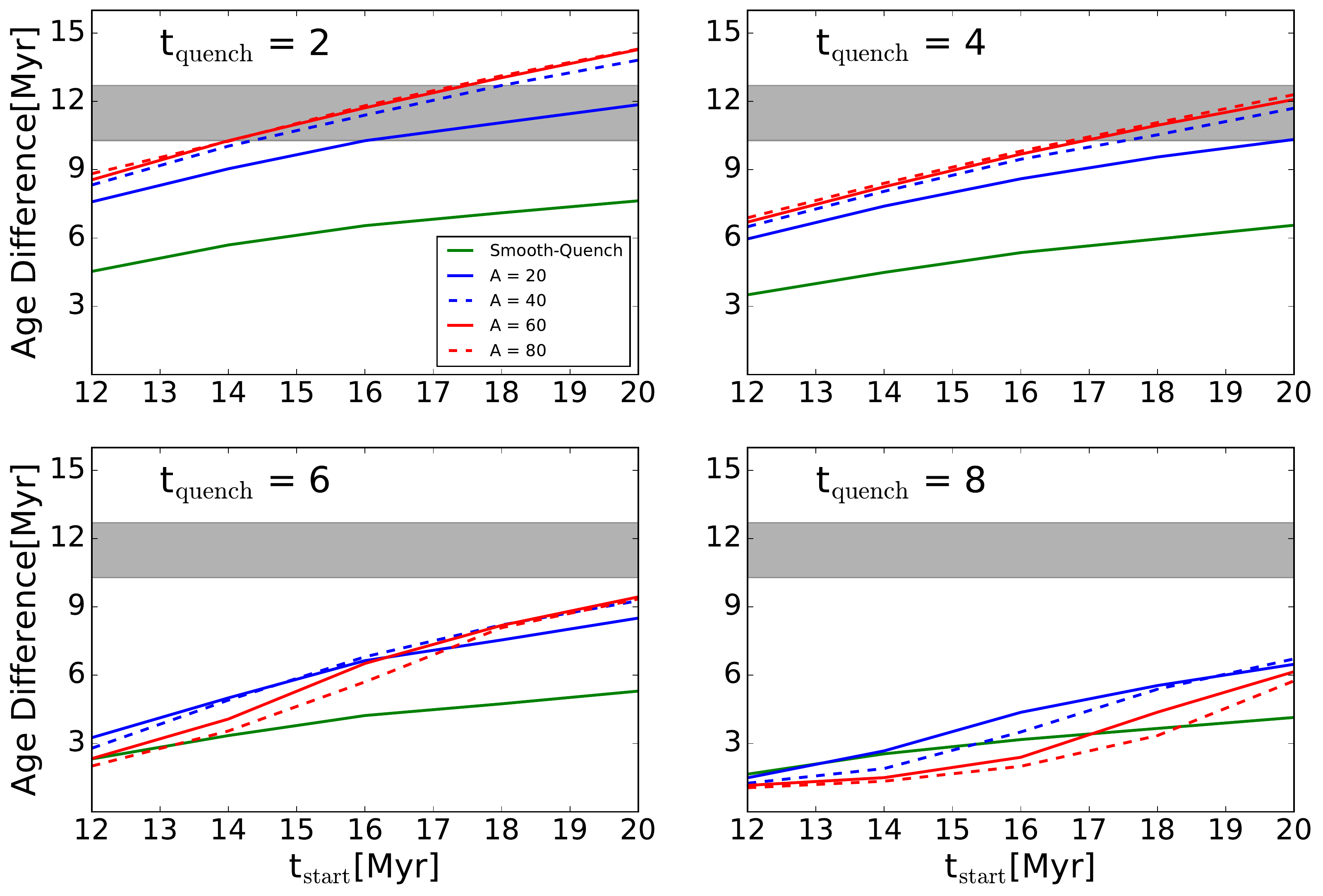}
\centering
\caption{The age difference between F5 and M0 stars versus $t_{\rm start}$ in our simulations for four different $t_{\rm quench}$ (four panels) and different $A$ values (color/styled lines as shown).  In general, the age differences are most signicant when star formation was quenched recently.  Age smooths out differences. A burst increases the age difference, although this effect saturates quickly.  The shaded region shows the measured difference between F5 and M0 stars in Upper Sco.}
\label{fig:Burst_A}
\end{figure*}


\subsubsection{Age Differences versus Photospheric Temperature}

When star formation occurs over some duration, hotter stars will have older ages than cooler stars.  Figure~\ref{fig:Burst_A_detail} shows the age versus temperature for a cluster that started forming stars 16 Myr ago and ceased forming stars 4 and 8 Myr ago.  For this plot, the start time of star formation is fixed at t$_{\rm start} = 16$ Myr, with the quench time $t_{\rm quench}$ fixed at 4 Myr (top panel) and $8$ Myr (bottom panel).  Figure~\ref{fig:Burst_A} shows these same differences, but parameterized by the age difference between F5 and M0 stars to show results from additional models. The time since the last star formation, $t_{\rm quench}$, is fixed in each of the panels. 

In the most basic Smooth-Quench model, the age differences between F5 and M0 stars are largest when star formation occurred recently and decrease with time.  The recent star formation maximizes the number of intermediate-mass stars that are still fully convective. If no star formation has occurred in 8 Myr, then the differences between ages of F5 and M0 stars would be $\sim 2-3$ Myr.  The duration of star formation ($t_{\rm start}-t_{\rm quench}$) increases these differences.  In the limit where $t_{\rm start}-t_{\rm quench}=0$, then there would be no difference in age between F5 and M0 stars since all star formation would occur instantaneously.

Moving to the burst model, a modest burst ($A=20$) increases the age difference by $\sim 4$ Myr.  Much larger bursts do not increase this effect significantly.  Again, as $t_{\rm quench}$ increases, the intermediate mass stars born during the burst will have enough time to evolve into radiative phases, thereby reducing any apparent age differences between F5 and M0 stars. Larger $A$-values will decrease the median ages hot and cold stars.  When $t_{\rm quench}$ is large (long duration since the burst), larger $A$ values will result in smaller age differences.


The age differences become larger as the duration of star formation, $t_{\rm start}-t_{\rm quench}$, increases.  For fixed $t_{\rm start}$ and $A$, larger $t_{\rm quench}$ will result in smaller age difference.  For a single burst, we should not expect a large age difference between stars of different spectral type. A lengthy star formation process ($t_{\rm start}- t_{\rm quench}$) is the most significant factor contributing to the spectral type-dependent age measurement.  



\begin{figure*}[!t]
\includegraphics[width = 13.5cm, height = 10.5cm]{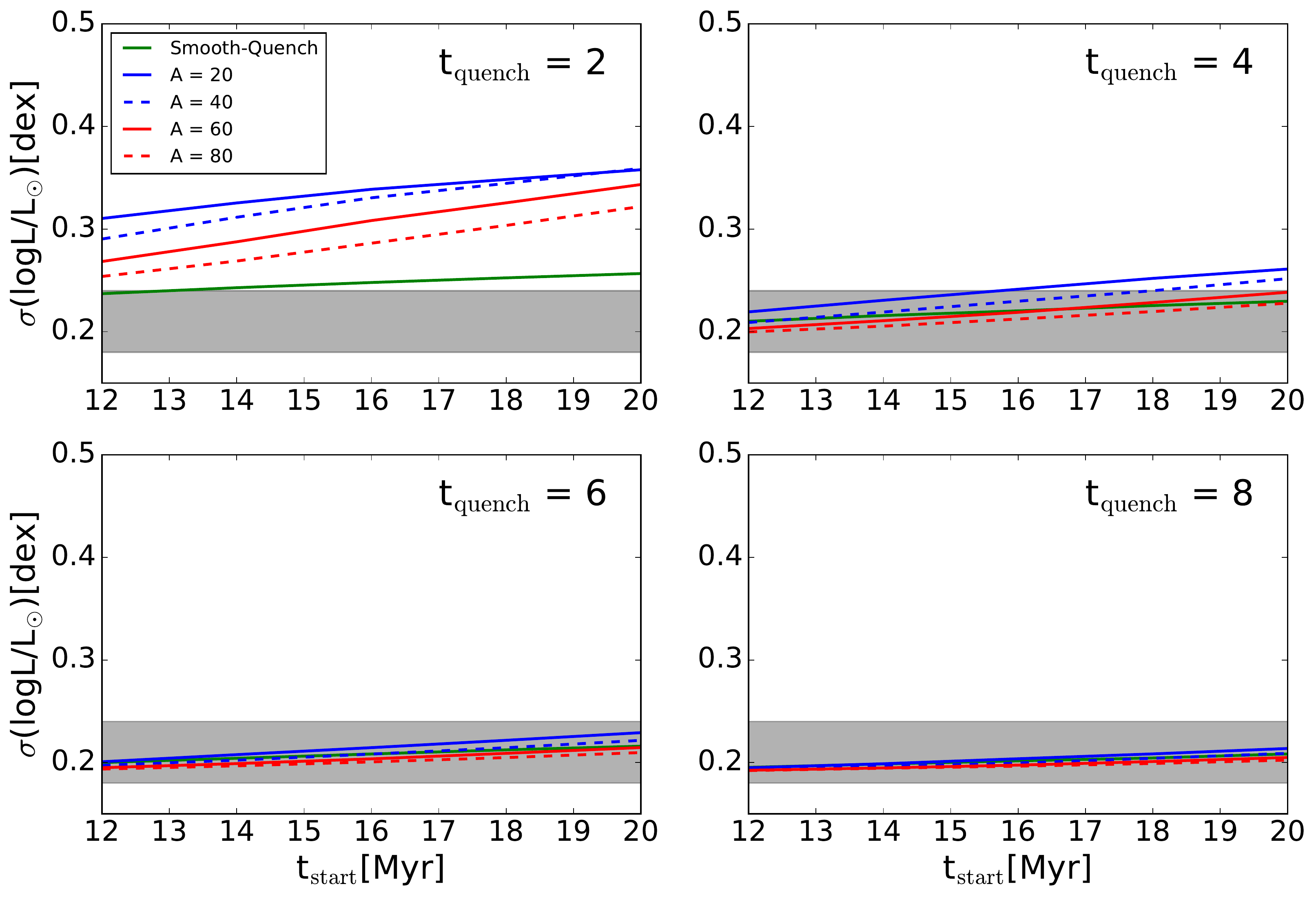}
\centering
\caption{Similar to Figure~\ref{fig:Burst_A}, but showing the luminosity spread of M0 stars instead of the age difference.  The shaded region shows the measured luminosity spread of M0 stars in Upper Sco.  The age spreads should be large if star formation was quenched recently and will gradually decrease with time.
\\}
\label{fig:Burst_lum_A}
\end{figure*}


\subsubsection{Luminosity Spread}
Any age spread during star formation will also lead to a spread in observed luminosities.  
In this subsection, we describe how different parameters in Smooth-Quench and Smooth-Burst models affect the luminosity spread of M0 type stars.  Earlier spectral types are affected by stellar evolution and are a poor probe of age spread with additional analysis, while M0 stars are fully convective and contract along the Hayashi track for the duration of the first 20 Myr of evolution simulated here.

Figure~\ref{fig:Burst_lum_A} show the effects of $A$ and $t_{\rm start}$ on the luminosity spread.  The luminosity spreads are always largest when star formation is recent because the luminosity of an M0 star decreases with $\log$ age, while we are measuring age spread in linear space.  
In the Smooth-Quench model, for example the age spread (standard deviation of all ages) will be $\sim 0.29 \times (t_{\rm start}-t_{\rm quench})$.  While this age spread is constant, the luminosity spread decreases with age.

All other factors are second-order effects in the luminosity spread.  A modest burst ($A=20$) will increase the luminosity spread.  In a large burst ($A=80$), the number of cool stars born during the burst overwhelm the number of stars formed during the smooth star formation, thereby reducing the age spread.



\section{APPLYING STAR FORMATION HISTORIES TO UPPER SCO}
The Upper Sco region covers a projected distance of 30 pc on the sky, with a sound crossing time of $\sim 85$ Myr \citep[see discussion in][]{slesnick08}.  The lack of any known Class 0 or I stars indicates that all star formation within Upper Sco has ceased.  \citet{degeus92} and \citet{preibisch99} suggest that star formation in Upper Sco may have been triggered by a supernova-driven shock wave from a massive star in the parent Sco-Cen OB Association \citep[see also][]{preibisch07,slesnick08}.  In this context, smooth star formation could occur over a long interval, followed by a supernova that triggers a burst and subsequent quenching of star formation. 

\begin{figure}[!t]
\includegraphics[width = 8cm, height = 13.333cm]{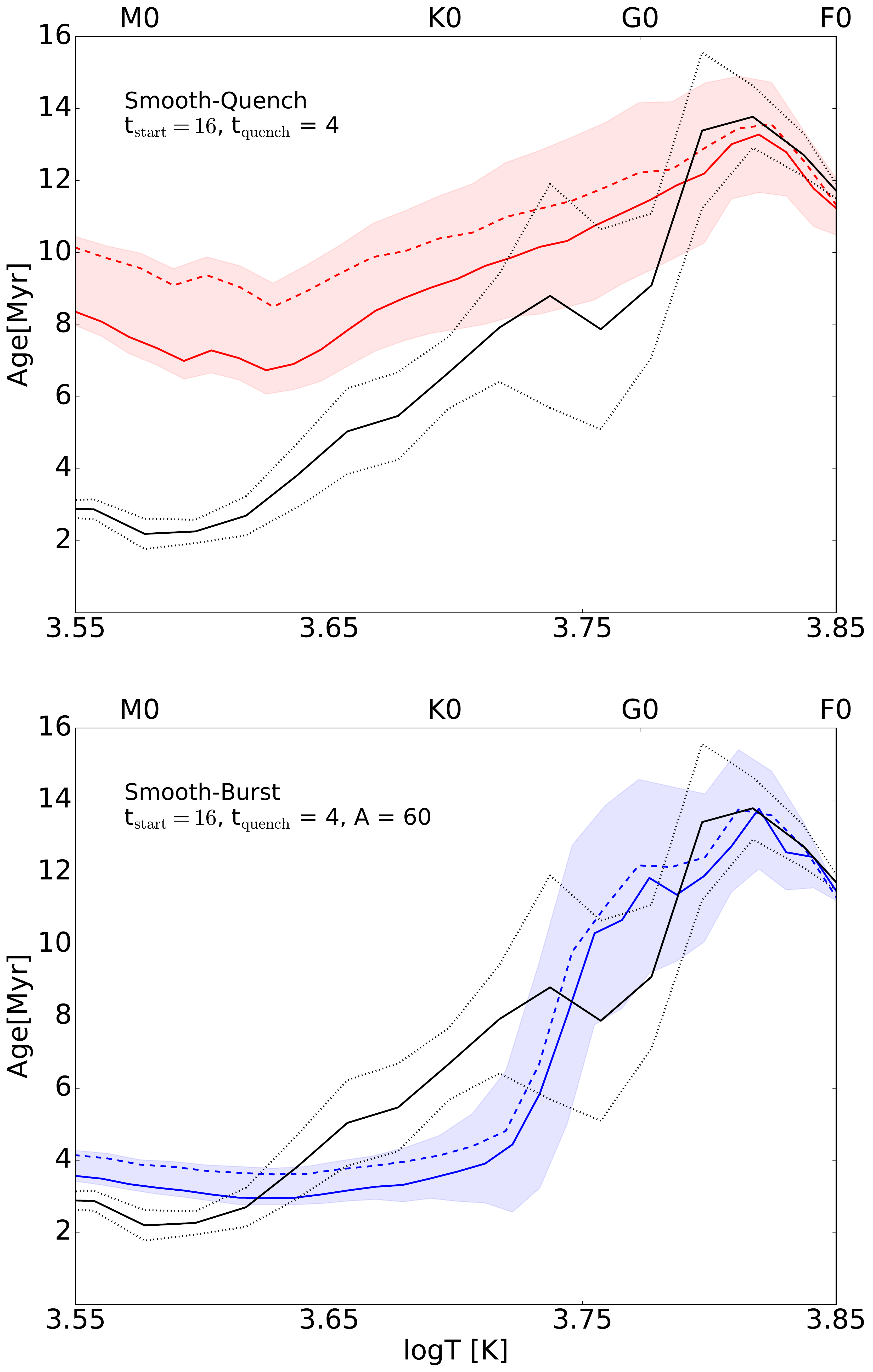}
\centering
\caption{The inferred age versus temperature of stars in Upper Sco (black, standard deviation in gray dotted lines), compared with results from the Smooth-Quench (top) model, with $t_{\rm start}=16$ Myr and $t_{\rm quench}=4$ Myr, and the Smooth-Burst model (bottom) with $A=60$, $t_{\rm start} = 16$ Myr, and $t_{\rm quench} = 4$ Myr.  The shaded regions as the $1 \sigma$ standard deviation from 1000 simulations.   The dashed line shows the same results, but when all binaries are resolved.}
\label{fig:Result_Observation}
\end{figure}

\begin{figure}[!t]
\includegraphics[width = 8cm, height = 6.222cm]{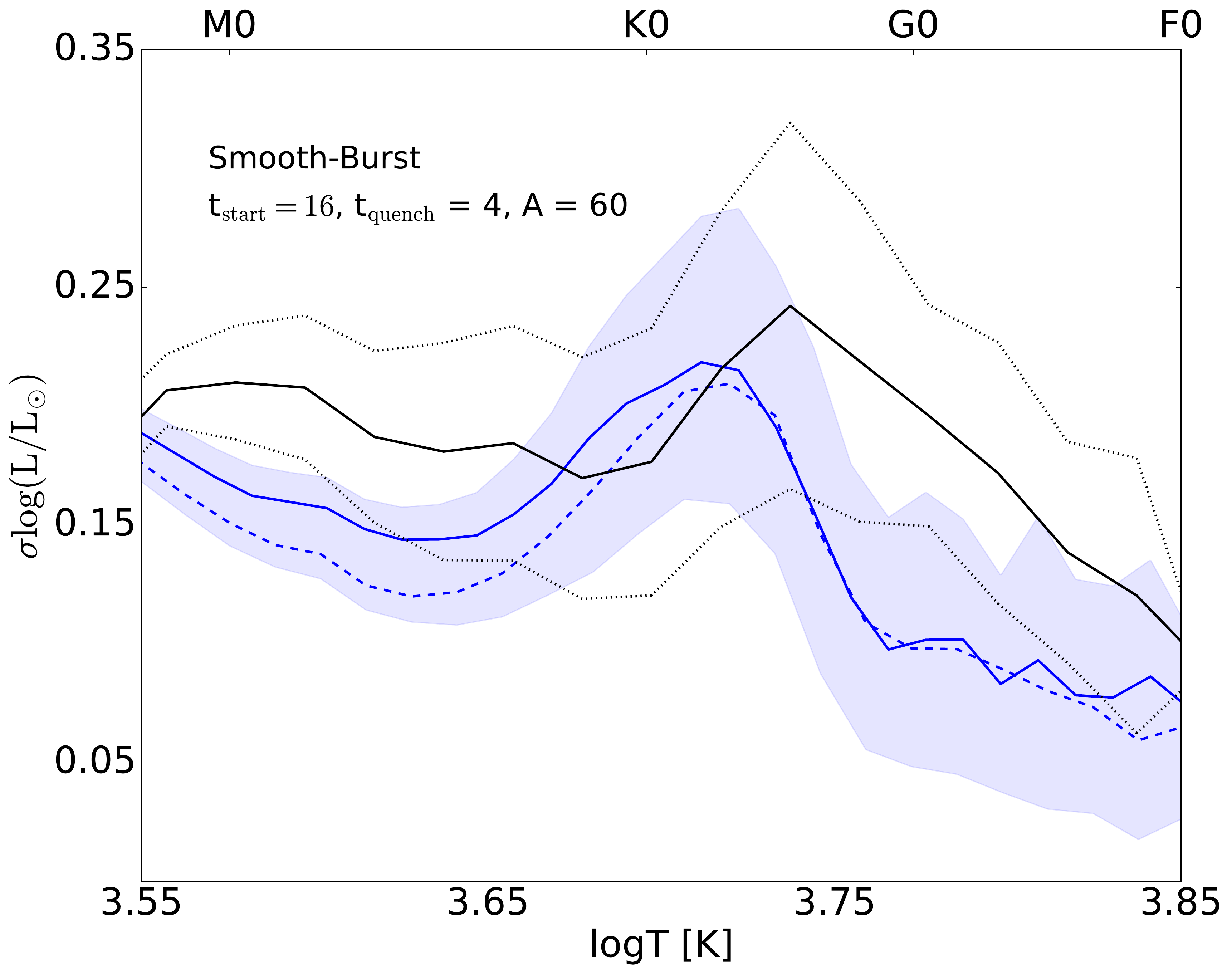}
\centering
\caption{The luminosity spread of Upper Sco in each temperature bin (solid black line, with gray shaded region as the $1\sigma$ uncertainty) versus results from the Smooth-Burst model (blue solid line, shaded region as the standard deviation from 1000 simulations, parameters as in Figure~\ref{fig:Result_Observation}).   The dashed line shows the same results, but when all binaries are resolved.}
\label{fig:agespread}
\end{figure}

\subsection{Comparing Synthetic Populations to the Known Upper Sco Members}

Our simulations of the star formation history of a cluster correspond to smooth star formation with no trigger (the Smooth-Quench model) and smooth star formation followed by a supernova trigger (Smooth-Burst model).  Figure ~\ref{fig:Result_Observation} compares of the results of these simulations to the stellar population in Upper Sco for the Smooth-Burst model with parameters $t_{\rm start} = 16$ Myr, $t_{\rm quench} = 4$ Myr and $A = 60$, and for the Smooth-Quench model with $t_{\rm start} = 16$ Myr and $t_{\rm quench} = 4$ Myr.  

The Smooth-Quench model exhibits a temperature-dependent age measurement up to $\sim 6$ Myr between KM stars and F stars, thereby accounting for half of the $\sim 12$ Myr discrepancy (see Table \ref{tab:ages.tab}).  This simplest model demonstrates that an age spread of more than a few Myr in a young ($<15$ Myr) population will lead to a significant temperature dependence in age estimates. This qualitative result is required by the evolution of intermediate mass stars from the Hayashi track to the Henyey track.   Cool stars are young and have a median age that changes slowly.  For stars with $\log T_{\rm eff}/K=3.6-3.7$, the median age increases rapidly.  At temperatures higher than $\log T_{\rm eff}/K>3.7$, stars tend to be older and have median ages that change slowly.  These simulation results are consistent with the age spreads of 4 -- 5 Myr over a relatively large temperature range of cool stars \citep[e.g.][]{herczeg15}, and with the $\sim 3$ Myr age spread for G type and F type stars \citep{pecaut12}.  

Figure~\ref{fig:Result_Observation} shows that the Smooth-Burst model may explain entirely the empirical temperature-age relationship of Upper Sco.
The burst increases the population of cool stars at the end of the star formation process, thereby maximizing the possible differences in ages between cool and hot stars. 

Some of the age difference between K-M type and F type stars may be attributed to the incomplete physics in the calculations of pre-main sequence evolution of low-mass stars \citep[e.g.][]{feiden16}, although these explanations may introduce problems into other clusters.  However, 
the age difference between G type stars ($9\pm3$ Myr) and F type stars ($13\pm1$ Myr), as calculated by \citet{pecaut12}, are not easily explained by the same changes that are invoked to move the K-M type stars to older ages.  The Smooth-Burst model recovers this age difference between G and F type stars without invoking any shortcomings in stellar evolution models or in empirical temperature and luminosity measurements.

Figure~\ref{fig:agespread} compares the simulated temperature-dependent luminosity spread to the measured luminosity spread within each temperature bin.  The luminosity spread in the simulations and observations are consistent at the level of 1$\sigma$. The luminosity spread in the simulation peaks at slightly cooler temperatures than the observed luminosity spread, and also yields luminosity spreads that are systemically smaller than the measured spreads.  Both of these uncertainties may be caused by observational errors, including sample incompleteness and mistaken membership, and also by errors in the stellar evolution models.  A full evaluation of the star formation history will require a more complete membership list that is free of any biases with spectral type.

\begin{figure*}[!th]
\includegraphics[width = 15cm, height = 10cm]{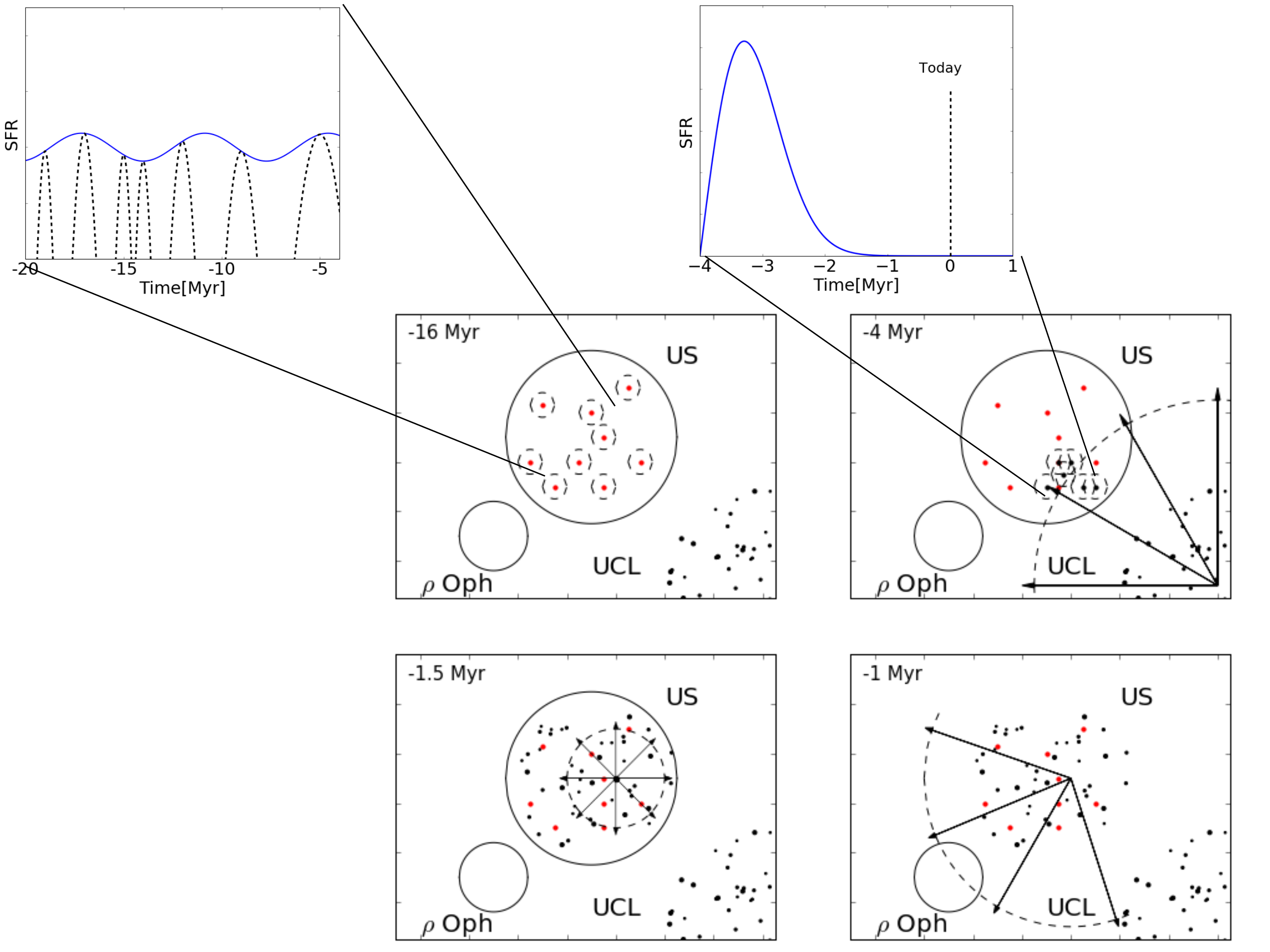}
\centering
\caption{A revised star formation history of Scorpius-Centaurus association based on the work of \citet{preibisch99}. Circle solid areas are molecular cloud and dashed circles are shock wave front. Stars are shown as dots and specially red dots are early-formed stars in US. See text for more details.}
\label{fig:Schematic_SFH}
\end{figure*}

\subsection{Revised Star Formation History of Upper Sco}

The star formation in Upper Sco may have been triggered $\sim 5$ Myr ago by the shock wave from a supernova explosion in Upper Centaurus Lupus \citep{degeus92,preibisch99}.  This star formation may have then been halted by a shock wave from another supernova that exploded in Upper Sco, destroying the molecular cloud. About 1 Myr ago, this shock wave would have arrived at the $\rho$ Oph molecular cloud and triggered star formation.  The scenario of the shock wave from Upper Centaurus Lupus triggering star formation in Upper Sco is challenged by the $\sim 11$ Myr old age of Upper Sco, as measured from intermediate- and high-mass stars \citep{pecaut12,mamajek13}, but is consistent with the younger age estimated for low-mass stars.  The conflict in the age measurements of different spectral types in the same cluster is an impediment toward our understanding of star formation history. 

In this work, we find that when a lengthy star formation is introduced before the trigger event, stars of different spectral types give different age estimations. In this context, we speculate about a plausible revision to the sequentially trigger star formation proposed by \citet{preibisch99}, with a possible scenario outlined as follows (see also Figure~\ref{fig:Schematic_SFH}): the star formation in Upper Sco started about $16$ Myr ago with a constant star formation rate with time. About 12 Myr ago, the most massive star in Upper Centaurus Lupus exploded, creating a shock wave that arrived at Upper Sco 5 Myr ago, triggering a burst of star formation. About 3 Myr later, the most massive star in Upper Sco may have destroyed the parent molecular cloud quenched star formation.  This star would need to have formed early during the initial constant phase of star formation, because the formation of a $20$ M$_\odot$ star, which explodes in $\sim 10$ Myr, is far more likely than a $60$ M$_\odot$ star, which explodes in $\sim 3$ Myr \citep{ekstrom12}.   The shock wave from this supernova explosion may have then arrived at $\rho$ Oph, triggering star formation there.  Both supernova and hot star winds are expected to have this type of positive and negative feedback from simulations \citep[e.g.][]{dale12} and also observations of clusters in massive star forming complexes \citep[e.g.][]{bik12,rivera13,jose17}.

Compared with the scenario proposed by \citet{preibisch99}, the similar history includes weak star formation with a long duration before the trigger event.  This constant star formation may be related to that of Upper Centaurus Lupus and Lower Centaurus Crux, which have older average ages than Upper Sco \citep[e.g.][]{pecaut16}.
The physics behind a lengthy star formation may be related to the cooling down of molecular cloud. The depletion time, i.e. the lag time between the formation of molecular time and the start time of the burst of star formation, is an important timescale for the understanding of molecular cloud evolution. As a molecular cloud cools, star formation can occur in over-dense regions; the star formation is local so the star formation rate would be low.  The over-dense regions may be distributed randomly over the molecular cloud, with small bursts that appears to be a constant star formation with time. Although the estimation on depletion time requires larger samples and more detailed simulations, the temperature-dependent age estimate may provide a lower limit on past star formation.

\section{Discussion and Conclusions}

This paper investigates the effects of stellar evolution in analyses of ages and histories of young stellar clusters. 
Ages of young associations are often estimated in specific spectral type bins \citep[e.g.][]{pecaut12,herczeg15} or estimated by fitting empirical isochrones to clusters \citep{bell15}, with membership lists that are incomplete and may be biased to hotter or cooler stars for different regions.  However, if star formation proceeds over millions of years, then the different pre-main sequence tracks of intermediate-mass and low-mass stars will lead directly to a spectral type (temperature) dependence in the star formation history.  These differences arise when ages are measured in spectral type (or temperature) bins rather than in mass, which is a pragmatic approach because mass is not a measurable quantity.  

In this context, in Upper Sco about half of the age discrepancy between F and KM stars may be attributed to the star formation history.  A star formation burst at the end of the star formation epoch will maximize these differences and could entirely explain the locus of stars in the HR diagram of Upper Sco.   These simulations also explain the differences in median age between F and G stars in Upper Sco, where any age discrepancy is less likely to be introduced by problems in models of pre-main sequence evolution.  The accelerated star formation scenario of \citet{palla00} would be intermediate between the two scenarios modeled here.  Isochrone fitting may also suffer from related problems when the known members of different clusters have a different distribution in spectral type or mass.  Although the use of Dotter models maximizes differences between intermediate-mass and low-mass stars, other models yield qualitatively similar results.  Very low-mass star and brown dwarfs in young associations have much younger ages than assessed from the Li Depletion Boundary \citep[e.g.][]{malo14}, which cannot be explained by age spreads and is instead likely caused by errors in evolutionary tracks.

These effects are only present if star formation occurs over some duration of time, 
and are most significant if the timescale for star formation in the cluster are similar to the timescales for intermediate mass stars to transition from convective to radiative interiors.  These timescales correspond to the expected age and possible age spread of Upper Sco.  In some cases, most star formation may occur in a single burst, as may be the case for the Orion Nebula Cluster \citep{reggiani11} and for regions of triggered regions of star formation within more massive clusters \citep[e.g.][]{rivera13,jose17}, although even in these cases some past of star formation would affect age measurements for hotter stars without measurable differences for cooler stars.  If star formation occurs over an extended timescale of $\sim 5-10$ Myr, then the early-to-mid M-dwarfs will provide the most robust historical record of star formation rate versus time, including average age of stars in the cluster and the luminosity/age spread.  This mode of star formation may be quite common, based on inferences of age spreads in high-mass star forming regions \citep[e.g.][]{bik12,kiminki15,jose16}.  The early spectral types will provide a record of the earliest epoch of star formation \citep[e.g.][]{roman15,kiminki15,wu16} but will be biased against the most recent epochs of star formation.

As a significant caveat to the application of the simulations to Upper Sco, observational uncertainties in the effective temperature may be especially large.  A dedicated study of a single T Tauri star indicates that spots may radically change effective temperatures of low-mass young stars \citep{gully17}.  Spots are likely common and large on low-mass stars through the $\sim 125$ Myr age of the Pleiades  \citep{fang16}, which may inflate radii \citep{somers16}.   Some of the age difference between cool and hot stars may be driven by this observational uncertainty and deficiencies in the physics of the pre-main sequence models.  Updated stellar models of \citet{somers15} include spots but are not yet available as a grid and may be challenging to apply to large datasets with stars that have a wide range of spot properties. 

Pre-main sequence evolution is still an unsolved problem, with unsettled differences between different models and between models and observations.  
The age discrepancies between stars of different temperatures in Upper Sco have helped to motivate a new generation of models.  Our results here suggest that, while such re-evaluation is important, it is unclear whether these improvements should be judged by comparing coeval isochrones to the population of Upper Sco.  A complete census of members in Upper Sco, without any bias versus spectral type, will be necessary to test whether the age discrepancy versus spectral type is driven primarily by an age spread or is the result of large errors in pre-main sequence evolution.

\vspace{10mm}

\section{Appendix A: Applying magnetic Feiden models to young star-forming regions}

The \citet{feiden16} evolutionary models introduce magnetic fields in the stellar interior, which reduce the efficiency of convection and thereby slows contraction rates.  The available models include large fluctuations at young ages ($<10$ Myr), before the results settle into a stable configuration.  We adapt these models by assuming that the evolution at slightly older ages can be extrapolated to young ages with linear fits.  This extrapolation then allows these evolutionary tracks to be applied to young associations. 


Table~\ref{tab:feidenages} applies these models to several associations to obtain a consistent set of ages for these tracks.  The known members of each cluster were fit with an empirical isochrone by \citet{herczeg15}, which was then scaled to each region to obtain the expected median luminosities for stars with temperatures of 4200 K, 3800 K, and 3400 K.  In this appendix, the ages of 4200 K and 3400 K stars are calculated here using the magnetic Feiden tracks, which were not yet available in 2015.  Further details in this simplistic approach can be found in \citet{herczeg15}.

In the magnetic Feiden models, the inferred ages of a 4200 K low mass star are 40 Myr in the Beta Pic Moving Group and 50 Myr in the Tuc-Hor Association,  older than the respective Lithium depletion boundary ages of $\sim 24$ and $40$ Myr \citep{binks14,malo14,kraus14}.   Stars with spectral types later than M3 have much younger inferred ages for all regions, including Upper Sco, because of discrepancies between observations and predictions from pre-main sequence evolution \citep[see Discussion in][]{herczeg15}.


\begin{table}[!t]
\label{tab:feidenages}
\caption{Ages of Nearby Associations with Feiden magnetic models}
\begin{tabular}{ccccc}
Region & $\log L_{4200}/L_\odot$ & Age & $\log L_{3400}/L_\odot$ & Age\\
\hline
Upper Sco & $-0.25$ & $10.5$ & $-0.86$ & $6.2$\\
$\epsilon$ Cha & $-0.28$ & $11.6$ & $-0.89$ & $6.7$\\
$\eta$ Cha & $-0.32$ & $13.5$ & $-0.93$ & $7.6$\\
TWA &  $-0.40$ & 18  & $-1.01$ & $9.7$\\
BPMG & $-0.65$ & 40 & $-1.25$ & $21$\\
Tuc-Hor & $-0.72$ & 50 & $-1.33$ & $25$\\
\hline
\end{tabular}
\end{table}

\acknowledgements

We thank the anonymous referee for constructive comments, which helped improve the quality and clarity of the paper.  We also thank Adam Kraus, Michael Ireland, and Jinyi Shangguan for valuable discussions, and Garrett Somers and Greg Feiden for suggestions on the manuscript.  QF and GJH are supported by general grant 11473005 awarded by the National Science Foundation of China and by a course development grant awarded by Peking University.  

This work has made use of data from many facilities, including the Wide-field Infrared Survey Explorer, which is a joint project of the University of California, Los Angeles, and the Jet Propulsion Laboratory/California Institute of Technology, funded by the National Aeronautics and Space Administration; the Two Micron All Sky Survey, which is a joint project of the University of Massachusetts and the Infrared Processing and Analysis Center/California Institute of Technology, funded by the National Aeronautics and Space Administration and the National Science Foundation; the European Space Agency (ESA) mission Gaia, processed by the Gaia Data Processing and Analysis Consortium, funded by the institutions participating in the Gaia Multilateral Agreement, and the VizieR catalogue access tool, CDS, Strasbourg, France.

\bibliographystyle{apj}
\bibliography{ms_final}

\begin{thebibliography}{76}
\expandafter\ifx\csname natexlab\endcsname\relax\def\natexlab#1{#1}\fi
\expandafter\ifx\csname bibnamefont\endcsname\relax
  \def\bibnamefont#1{#1}\fi
\expandafter\ifx\csname bibfnamefont\endcsname\relax
  \def\bibfnamefont#1{#1}\fi
\expandafter\ifx\csname citenamefont\endcsname\relax
  \def\citenamefont#1{#1}\fi
\expandafter\ifx\csname url\endcsname\relax
  \def\url#1{\texttt{#1}}\fi
\expandafter\ifx\csname urlprefix\endcsname\relax\def\urlprefix{URL }\fi
\providecommand{\bibinfo}[2]{#2}
\providecommand{\eprint}[2][]{\url{#2}}

\bibitem[{\citenamefont{{Soderblom} et~al.}(2014)\citenamefont{{Soderblom},
  {Hillenbrand}, {Jeffries}, {Mamajek}, and {Naylor}}}]{soderblom14}
\bibinfo{author}{\bibfnamefont{D.~R.} \bibnamefont{{Soderblom}}},
  \bibinfo{author}{\bibfnamefont{L.~A.} \bibnamefont{{Hillenbrand}}},
  \bibinfo{author}{\bibfnamefont{R.~D.} \bibnamefont{{Jeffries}}},
  \bibinfo{author}{\bibfnamefont{E.~E.} \bibnamefont{{Mamajek}}},
  \bibnamefont{and} \bibinfo{author}{\bibfnamefont{T.}~\bibnamefont{{Naylor}}},
  \bibinfo{journal}{Protostars and Planets VI} pp. \bibinfo{pages}{219--241}
  (\bibinfo{year}{2014}), \eprint{1311.7024}.

\bibitem[{\citenamefont{{Hartmann} et~al.}(2016)\citenamefont{{Hartmann},
  {Herczeg}, and {Calvet}}}]{hartmann16}
\bibinfo{author}{\bibfnamefont{L.}~\bibnamefont{{Hartmann}}},
  \bibinfo{author}{\bibfnamefont{G.}~\bibnamefont{{Herczeg}}},
  \bibnamefont{and} \bibinfo{author}{\bibfnamefont{N.}~\bibnamefont{{Calvet}}},
  \bibinfo{journal}{\araa} \textbf{\bibinfo{volume}{54}}, \bibinfo{pages}{135}
  (\bibinfo{year}{2016}).

\bibitem[{\citenamefont{{de Geus} et~al.}(1989)\citenamefont{{de Geus}, {de
  Zeeuw}, and {Lub}}}]{degeus89}
\bibinfo{author}{\bibfnamefont{E.~J.} \bibnamefont{{de Geus}}},
  \bibinfo{author}{\bibfnamefont{P.~T.} \bibnamefont{{de Zeeuw}}},
  \bibnamefont{and} \bibinfo{author}{\bibfnamefont{J.}~\bibnamefont{{Lub}}},
  \bibinfo{journal}{\aap} \textbf{\bibinfo{volume}{216}}, \bibinfo{pages}{44}
  (\bibinfo{year}{1989}).

\bibitem[{\citenamefont{{de Zeeuw} et~al.}(1999)\citenamefont{{de Zeeuw},
  {Hoogerwerf}, {de Bruijne}, {Brown}, and {Blaauw}}}]{dezeeuw99}
\bibinfo{author}{\bibfnamefont{P.~T.} \bibnamefont{{de Zeeuw}}},
  \bibinfo{author}{\bibfnamefont{R.}~\bibnamefont{{Hoogerwerf}}},
  \bibinfo{author}{\bibfnamefont{J.~H.~J.} \bibnamefont{{de Bruijne}}},
  \bibinfo{author}{\bibfnamefont{A.~G.~A.} \bibnamefont{{Brown}}},
  \bibnamefont{and} \bibinfo{author}{\bibfnamefont{A.}~\bibnamefont{{Blaauw}}},
  \bibinfo{journal}{\aj} \textbf{\bibinfo{volume}{117}}, \bibinfo{pages}{354}
  (\bibinfo{year}{1999}), \eprint{astro-ph/9809227}.

\bibitem[{\citenamefont{{Preibisch} et~al.}(2002)\citenamefont{{Preibisch},
  {Brown}, {Bridges}, {Guenther}, and {Zinnecker}}}]{preibisch02}
\bibinfo{author}{\bibfnamefont{T.}~\bibnamefont{{Preibisch}}},
  \bibinfo{author}{\bibfnamefont{A.~G.~A.} \bibnamefont{{Brown}}},
  \bibinfo{author}{\bibfnamefont{T.}~\bibnamefont{{Bridges}}},
  \bibinfo{author}{\bibfnamefont{E.}~\bibnamefont{{Guenther}}},
  \bibnamefont{and}
  \bibinfo{author}{\bibfnamefont{H.}~\bibnamefont{{Zinnecker}}},
  \bibinfo{journal}{\aj} \textbf{\bibinfo{volume}{124}}, \bibinfo{pages}{404}
  (\bibinfo{year}{2002}).

\bibitem[{\citenamefont{{Slesnick} et~al.}(2008)\citenamefont{{Slesnick},
  {Hillenbrand}, and {Carpenter}}}]{slesnick08}
\bibinfo{author}{\bibfnamefont{C.~L.} \bibnamefont{{Slesnick}}},
  \bibinfo{author}{\bibfnamefont{L.~A.} \bibnamefont{{Hillenbrand}}},
  \bibnamefont{and} \bibinfo{author}{\bibfnamefont{J.~M.}
  \bibnamefont{{Carpenter}}}, \bibinfo{journal}{\apj}
  \textbf{\bibinfo{volume}{688}}, \bibinfo{eid}{377-397}
  (\bibinfo{year}{2008}), \eprint{0809.1436}.

\bibitem[{\citenamefont{{Rizzuto} et~al.}(2015)\citenamefont{{Rizzuto},
  {Ireland}, and {Kraus}}}]{rizzuto15}
\bibinfo{author}{\bibfnamefont{A.~C.} \bibnamefont{{Rizzuto}}},
  \bibinfo{author}{\bibfnamefont{M.~J.} \bibnamefont{{Ireland}}},
  \bibnamefont{and} \bibinfo{author}{\bibfnamefont{A.~L.}
  \bibnamefont{{Kraus}}}, \bibinfo{journal}{\mnras}
  \textbf{\bibinfo{volume}{448}}, \bibinfo{pages}{2737} (\bibinfo{year}{2015}),
  \eprint{1501.07270}.

\bibitem[{\citenamefont{{Pecaut} et~al.}(2012)\citenamefont{{Pecaut},
  {Mamajek}, and {Bubar}}}]{pecaut12}
\bibinfo{author}{\bibfnamefont{M.~J.} \bibnamefont{{Pecaut}}},
  \bibinfo{author}{\bibfnamefont{E.~E.} \bibnamefont{{Mamajek}}},
  \bibnamefont{and} \bibinfo{author}{\bibfnamefont{E.~J.}
  \bibnamefont{{Bubar}}}, \bibinfo{journal}{\apj}
  \textbf{\bibinfo{volume}{746}}, \bibinfo{eid}{154} (\bibinfo{year}{2012}),
  \eprint{1112.1695}.

\bibitem[{\citenamefont{{Herczeg} and {Hillenbrand}}(2015)}]{herczeg15}
\bibinfo{author}{\bibfnamefont{G.~J.} \bibnamefont{{Herczeg}}}
  \bibnamefont{and} \bibinfo{author}{\bibfnamefont{L.~A.}
  \bibnamefont{{Hillenbrand}}}, \bibinfo{journal}{\apj}
  \textbf{\bibinfo{volume}{808}}, \bibinfo{eid}{23} (\bibinfo{year}{2015}),
  \eprint{1505.06518}.

\bibitem[{\citenamefont{{Rizzuto} et~al.}(2016)\citenamefont{{Rizzuto},
  {Ireland}, {Dupuy}, and {Kraus}}}]{rizzuto16}
\bibinfo{author}{\bibfnamefont{A.~C.} \bibnamefont{{Rizzuto}}},
  \bibinfo{author}{\bibfnamefont{M.~J.} \bibnamefont{{Ireland}}},
  \bibinfo{author}{\bibfnamefont{T.~J.} \bibnamefont{{Dupuy}}},
  \bibnamefont{and} \bibinfo{author}{\bibfnamefont{A.~L.}
  \bibnamefont{{Kraus}}}, \bibinfo{journal}{\apj}
  \textbf{\bibinfo{volume}{817}}, \bibinfo{eid}{164} (\bibinfo{year}{2016}),
  \eprint{1512.05371}.

\bibitem[{\citenamefont{{Pecaut} and {Mamajek}}(2016)}]{pecaut16}
\bibinfo{author}{\bibfnamefont{M.~J.} \bibnamefont{{Pecaut}}} \bibnamefont{and}
  \bibinfo{author}{\bibfnamefont{E.~E.} \bibnamefont{{Mamajek}}},
  \bibinfo{journal}{\mnras} \textbf{\bibinfo{volume}{461}},
  \bibinfo{pages}{794} (\bibinfo{year}{2016}), \eprint{1605.08789}.

\bibitem[{\citenamefont{{Hillenbrand}}(1997)}]{hillenbrand97}
\bibinfo{author}{\bibfnamefont{L.~A.} \bibnamefont{{Hillenbrand}}},
  \bibinfo{journal}{\aj} \textbf{\bibinfo{volume}{113}}, \bibinfo{pages}{1733}
  (\bibinfo{year}{1997}).

\bibitem[{\citenamefont{{Hillenbrand} et~al.}(2008)\citenamefont{{Hillenbrand},
  {Bauermeister}, and {White}}}]{hillenbrand08}
\bibinfo{author}{\bibfnamefont{L.~A.} \bibnamefont{{Hillenbrand}}},
  \bibinfo{author}{\bibfnamefont{A.}~\bibnamefont{{Bauermeister}}},
  \bibnamefont{and} \bibinfo{author}{\bibfnamefont{R.~J.}
  \bibnamefont{{White}}}, in \emph{\bibinfo{booktitle}{14th Cambridge Workshop
  on Cool Stars, Stellar Systems, and the Sun}}, edited by
  \bibinfo{editor}{\bibfnamefont{G.}~\bibnamefont{{van Belle}}}
  (\bibinfo{year}{2008}), vol. \bibinfo{volume}{384} of
  \emph{\bibinfo{series}{Astronomical Society of the Pacific Conference
  Series}}, p. \bibinfo{pages}{200}, \eprint{astro-ph/0703642}.

\bibitem[{\citenamefont{{Naylor}}(2009)}]{naylor09}
\bibinfo{author}{\bibfnamefont{T.}~\bibnamefont{{Naylor}}},
  \bibinfo{journal}{\mnras} \textbf{\bibinfo{volume}{399}},
  \bibinfo{pages}{432} (\bibinfo{year}{2009}), \eprint{0907.2307}.

\bibitem[{\citenamefont{{Malo} et~al.}(2014)\citenamefont{{Malo}, {Doyon},
  {Feiden}, {Albert}, {Lafreni{\`e}re}, {Artigau}, {Gagn{\'e}}, and
  {Riedel}}}]{malo14}
\bibinfo{author}{\bibfnamefont{L.}~\bibnamefont{{Malo}}},
  \bibinfo{author}{\bibfnamefont{R.}~\bibnamefont{{Doyon}}},
  \bibinfo{author}{\bibfnamefont{G.~A.} \bibnamefont{{Feiden}}},
  \bibinfo{author}{\bibfnamefont{L.}~\bibnamefont{{Albert}}},
  \bibinfo{author}{\bibfnamefont{D.}~\bibnamefont{{Lafreni{\`e}re}}},
  \bibinfo{author}{\bibfnamefont{{\'E}.}~\bibnamefont{{Artigau}}},
  \bibinfo{author}{\bibfnamefont{J.}~\bibnamefont{{Gagn{\'e}}}},
  \bibnamefont{and} \bibinfo{author}{\bibfnamefont{A.}~\bibnamefont{{Riedel}}},
  \bibinfo{journal}{\apj} \textbf{\bibinfo{volume}{792}}, \bibinfo{eid}{37}
  (\bibinfo{year}{2014}), \eprint{1406.6750}.

\bibitem[{\citenamefont{{Bell} et~al.}(2015)\citenamefont{{Bell}, {Mamajek},
  and {Naylor}}}]{bell15}
\bibinfo{author}{\bibfnamefont{C.~P.~M.} \bibnamefont{{Bell}}},
  \bibinfo{author}{\bibfnamefont{E.~E.} \bibnamefont{{Mamajek}}},
  \bibnamefont{and} \bibinfo{author}{\bibfnamefont{T.}~\bibnamefont{{Naylor}}},
  \bibinfo{journal}{\mnras} \textbf{\bibinfo{volume}{454}},
  \bibinfo{pages}{593} (\bibinfo{year}{2015}), \eprint{1508.05955}.

\bibitem[{\citenamefont{{Feiden}}(2016)}]{feiden16}
\bibinfo{author}{\bibfnamefont{G.~A.} \bibnamefont{{Feiden}}},
  \bibinfo{journal}{ArXiv e-prints}  (\bibinfo{year}{2016}),
  \eprint{1604.08036}.

\bibitem[{\citenamefont{{MacDonald} and {Mullan}}(2017)}]{macdonald17}
\bibinfo{author}{\bibfnamefont{J.}~\bibnamefont{{MacDonald}}} \bibnamefont{and}
  \bibinfo{author}{\bibfnamefont{D.~J.} \bibnamefont{{Mullan}}},
  \bibinfo{journal}{\apj} \textbf{\bibinfo{volume}{834}}, \bibinfo{eid}{67}
  (\bibinfo{year}{2017}), \eprint{1608.02136}.

\bibitem[{\citenamefont{{Somers} and {Stassun}}(2017)}]{somers17}
\bibinfo{author}{\bibfnamefont{G.}~\bibnamefont{{Somers}}} \bibnamefont{and}
  \bibinfo{author}{\bibfnamefont{K.~G.} \bibnamefont{{Stassun}}},
  \bibinfo{journal}{\aj} \textbf{\bibinfo{volume}{153}}, \bibinfo{eid}{101}
  (\bibinfo{year}{2017}), \eprint{1609.04841}.

\bibitem[{\citenamefont{{Somers} and {Pinsonneault}}(2015)}]{somers15}
\bibinfo{author}{\bibfnamefont{G.}~\bibnamefont{{Somers}}} \bibnamefont{and}
  \bibinfo{author}{\bibfnamefont{M.~H.} \bibnamefont{{Pinsonneault}}},
  \bibinfo{journal}{\apj} \textbf{\bibinfo{volume}{807}}, \bibinfo{eid}{174}
  (\bibinfo{year}{2015}), \eprint{1506.01393}.

\bibitem[{\citenamefont{{Baraffe} et~al.}(2015)\citenamefont{{Baraffe},
  {Homeier}, {Allard}, and {Chabrier}}}]{baraffe15}
\bibinfo{author}{\bibfnamefont{I.}~\bibnamefont{{Baraffe}}},
  \bibinfo{author}{\bibfnamefont{D.}~\bibnamefont{{Homeier}}},
  \bibinfo{author}{\bibfnamefont{F.}~\bibnamefont{{Allard}}}, \bibnamefont{and}
  \bibinfo{author}{\bibfnamefont{G.}~\bibnamefont{{Chabrier}}},
  \bibinfo{journal}{\aap} \textbf{\bibinfo{volume}{577}}, \bibinfo{eid}{A42}
  (\bibinfo{year}{2015}), \eprint{1503.04107}.

\bibitem[{\citenamefont{{Hartmann} et~al.}(1997)\citenamefont{{Hartmann},
  {Cassen}, and {Kenyon}}}]{hartmann97}
\bibinfo{author}{\bibfnamefont{L.}~\bibnamefont{{Hartmann}}},
  \bibinfo{author}{\bibfnamefont{P.}~\bibnamefont{{Cassen}}}, \bibnamefont{and}
  \bibinfo{author}{\bibfnamefont{S.~J.} \bibnamefont{{Kenyon}}},
  \bibinfo{journal}{\apj} \textbf{\bibinfo{volume}{475}}, \bibinfo{pages}{770}
  (\bibinfo{year}{1997}).

\bibitem[{\citenamefont{{Baraffe} et~al.}(2016)\citenamefont{{Baraffe},
  {Elbakyan}, {Vorobyov}, and {Chabrier}}}]{baraffe16}
\bibinfo{author}{\bibfnamefont{I.}~\bibnamefont{{Baraffe}}},
  \bibinfo{author}{\bibfnamefont{V.~G.} \bibnamefont{{Elbakyan}}},
  \bibinfo{author}{\bibfnamefont{E.~I.} \bibnamefont{{Vorobyov}}},
  \bibnamefont{and}
  \bibinfo{author}{\bibfnamefont{G.}~\bibnamefont{{Chabrier}}},
  \bibinfo{journal}{ArXiv e-prints}  (\bibinfo{year}{2016}),
  \eprint{1608.07428}.

\bibitem[{\citenamefont{{Bell} et~al.}(2013)\citenamefont{{Bell}, {Naylor},
  {Mayne}, {Jeffries}, and {Littlefair}}}]{bell13}
\bibinfo{author}{\bibfnamefont{C.~P.~M.} \bibnamefont{{Bell}}},
  \bibinfo{author}{\bibfnamefont{T.}~\bibnamefont{{Naylor}}},
  \bibinfo{author}{\bibfnamefont{N.~J.} \bibnamefont{{Mayne}}},
  \bibinfo{author}{\bibfnamefont{R.~D.} \bibnamefont{{Jeffries}}},
  \bibnamefont{and} \bibinfo{author}{\bibfnamefont{S.~P.}
  \bibnamefont{{Littlefair}}}, \bibinfo{journal}{\mnras}
  \textbf{\bibinfo{volume}{434}}, \bibinfo{pages}{806} (\bibinfo{year}{2013}),
  \eprint{1306.3237}.

\bibitem[{\citenamefont{{Chen} et~al.}(2014)\citenamefont{{Chen}, {Girardi},
  {Bressan}, {Marigo}, {Barbieri}, and {Kong}}}]{chen14}
\bibinfo{author}{\bibfnamefont{Y.}~\bibnamefont{{Chen}}},
  \bibinfo{author}{\bibfnamefont{L.}~\bibnamefont{{Girardi}}},
  \bibinfo{author}{\bibfnamefont{A.}~\bibnamefont{{Bressan}}},
  \bibinfo{author}{\bibfnamefont{P.}~\bibnamefont{{Marigo}}},
  \bibinfo{author}{\bibfnamefont{M.}~\bibnamefont{{Barbieri}}},
  \bibnamefont{and} \bibinfo{author}{\bibfnamefont{X.}~\bibnamefont{{Kong}}},
  \bibinfo{journal}{\mnras} \textbf{\bibinfo{volume}{444}},
  \bibinfo{pages}{2525} (\bibinfo{year}{2014}), \eprint{1409.0322}.

\bibitem[{\citenamefont{{Hayashi}}(1961)}]{hayashi61}
\bibinfo{author}{\bibfnamefont{C.}~\bibnamefont{{Hayashi}}},
  \bibinfo{journal}{\pasj} \textbf{\bibinfo{volume}{13}}
  (\bibinfo{year}{1961}).

\bibitem[{\citenamefont{{Henyey} et~al.}(1955)\citenamefont{{Henyey},
  {Lelevier}, and {Lev{\'e}e}}}]{henyey55}
\bibinfo{author}{\bibfnamefont{L.~G.} \bibnamefont{{Henyey}}},
  \bibinfo{author}{\bibfnamefont{R.}~\bibnamefont{{Lelevier}}},
  \bibnamefont{and} \bibinfo{author}{\bibfnamefont{R.~D.}
  \bibnamefont{{Lev{\'e}e}}}, \bibinfo{journal}{\pasp}
  \textbf{\bibinfo{volume}{67}}, \bibinfo{pages}{154} (\bibinfo{year}{1955}).

\bibitem[{\citenamefont{{Hayashi} and {Nakano}}(1965)}]{hayashi65}
\bibinfo{author}{\bibfnamefont{C.}~\bibnamefont{{Hayashi}}} \bibnamefont{and}
  \bibinfo{author}{\bibfnamefont{T.}~\bibnamefont{{Nakano}}},
  \bibinfo{journal}{Progress of Theoretical Physics}
  \textbf{\bibinfo{volume}{34}}, \bibinfo{pages}{754} (\bibinfo{year}{1965}).

\bibitem[{\citenamefont{{Evans} et~al.}(2009)\citenamefont{{Evans}, {Dunham},
  {J{\o}rgensen}, {Enoch}, {Mer{\'{\i}}n}, {van Dishoeck}, {Alcal{\'a}},
  {Myers}, {Stapelfeldt}, {Huard} et~al.}}]{evans09}
\bibinfo{author}{\bibfnamefont{N.~J.} \bibnamefont{{Evans}},
  \bibfnamefont{II}}, \bibinfo{author}{\bibfnamefont{M.~M.}
  \bibnamefont{{Dunham}}}, \bibinfo{author}{\bibfnamefont{J.~K.}
  \bibnamefont{{J{\o}rgensen}}}, \bibinfo{author}{\bibfnamefont{M.~L.}
  \bibnamefont{{Enoch}}},
  \bibinfo{author}{\bibfnamefont{B.}~\bibnamefont{{Mer{\'{\i}}n}}},
  \bibinfo{author}{\bibfnamefont{E.~F.} \bibnamefont{{van Dishoeck}}},
  \bibinfo{author}{\bibfnamefont{J.~M.} \bibnamefont{{Alcal{\'a}}}},
  \bibinfo{author}{\bibfnamefont{P.~C.} \bibnamefont{{Myers}}},
  \bibinfo{author}{\bibfnamefont{K.~R.} \bibnamefont{{Stapelfeldt}}},
  \bibinfo{author}{\bibfnamefont{T.~L.} \bibnamefont{{Huard}}},
  \bibnamefont{et~al.}, \bibinfo{journal}{\apjs}
  \textbf{\bibinfo{volume}{181}}, \bibinfo{eid}{321-350}
  (\bibinfo{year}{2009}), \eprint{0811.1059}.

\bibitem[{\citenamefont{{Kraus} et~al.}(2017)\citenamefont{{Kraus}, {Herczeg},
  {Rizzuto}, {Mann}, {Slesnick}, {Carpenter}, {Hillenbrand}, and
  {Mamajek}}}]{kraus17}
\bibinfo{author}{\bibfnamefont{A.~L.} \bibnamefont{{Kraus}}},
  \bibinfo{author}{\bibfnamefont{G.~J.} \bibnamefont{{Herczeg}}},
  \bibinfo{author}{\bibfnamefont{A.~C.} \bibnamefont{{Rizzuto}}},
  \bibinfo{author}{\bibfnamefont{A.~W.} \bibnamefont{{Mann}}},
  \bibinfo{author}{\bibfnamefont{C.~L.} \bibnamefont{{Slesnick}}},
  \bibinfo{author}{\bibfnamefont{J.~M.} \bibnamefont{{Carpenter}}},
  \bibinfo{author}{\bibfnamefont{L.~A.} \bibnamefont{{Hillenbrand}}},
  \bibnamefont{and} \bibinfo{author}{\bibfnamefont{E.~E.}
  \bibnamefont{{Mamajek}}}, \bibinfo{journal}{ArXiv e-prints}
  (\bibinfo{year}{2017}), \eprint{1702.04341}.

\bibitem[{\citenamefont{{Sacco} et~al.}(2017)\citenamefont{{Sacco}, {Spina},
  {Randich}, {Palla}, {Parker}, {Jeffries}, {Jackson}, {Meyer}, {Mapelli},
  {Lanzafame} et~al.}}]{sacco17}
\bibinfo{author}{\bibfnamefont{G.~G.} \bibnamefont{{Sacco}}},
  \bibinfo{author}{\bibfnamefont{L.}~\bibnamefont{{Spina}}},
  \bibinfo{author}{\bibfnamefont{S.}~\bibnamefont{{Randich}}},
  \bibinfo{author}{\bibfnamefont{F.}~\bibnamefont{{Palla}}},
  \bibinfo{author}{\bibfnamefont{R.~J.} \bibnamefont{{Parker}}},
  \bibinfo{author}{\bibfnamefont{R.~D.} \bibnamefont{{Jeffries}}},
  \bibinfo{author}{\bibfnamefont{R.}~\bibnamefont{{Jackson}}},
  \bibinfo{author}{\bibfnamefont{M.~R.} \bibnamefont{{Meyer}}},
  \bibinfo{author}{\bibfnamefont{M.}~\bibnamefont{{Mapelli}}},
  \bibinfo{author}{\bibfnamefont{A.~C.} \bibnamefont{{Lanzafame}}},
  \bibnamefont{et~al.}, \bibinfo{journal}{ArXiv e-prints}
  (\bibinfo{year}{2017}), \eprint{1701.03741}.

\bibitem[{\citenamefont{{Goudfrooij} et~al.}(2011)\citenamefont{{Goudfrooij},
  {Puzia}, {Chandar}, and {Kozhurina-Platais}}}]{goudfrooij11}
\bibinfo{author}{\bibfnamefont{P.}~\bibnamefont{{Goudfrooij}}},
  \bibinfo{author}{\bibfnamefont{T.~H.} \bibnamefont{{Puzia}}},
  \bibinfo{author}{\bibfnamefont{R.}~\bibnamefont{{Chandar}}},
  \bibnamefont{and}
  \bibinfo{author}{\bibfnamefont{V.}~\bibnamefont{{Kozhurina-Platais}}},
  \bibinfo{journal}{\apj} \textbf{\bibinfo{volume}{737}}, \bibinfo{eid}{4}
  (\bibinfo{year}{2011}), \eprint{1105.1317}.

\bibitem[{\citenamefont{{Preibisch} and {Zinnecker}}(2007)}]{preibisch07}
\bibinfo{author}{\bibfnamefont{T.}~\bibnamefont{{Preibisch}}} \bibnamefont{and}
  \bibinfo{author}{\bibfnamefont{H.}~\bibnamefont{{Zinnecker}}}, in
  \emph{\bibinfo{booktitle}{Triggered Star Formation in a Turbulent ISM}},
  edited by \bibinfo{editor}{\bibfnamefont{B.~G.} \bibnamefont{{Elmegreen}}}
  \bibnamefont{and} \bibinfo{editor}{\bibfnamefont{J.}~\bibnamefont{{Palous}}}
  (\bibinfo{year}{2007}), vol. \bibinfo{volume}{237} of
  \emph{\bibinfo{series}{IAU Symposium}}, pp. \bibinfo{pages}{270--277},
  \eprint{astro-ph/0610826}.

\bibitem[{\citenamefont{{Luhman} and {Mamajek}}(2012)}]{luhman12}
\bibinfo{author}{\bibfnamefont{K.~L.} \bibnamefont{{Luhman}}} \bibnamefont{and}
  \bibinfo{author}{\bibfnamefont{E.~E.} \bibnamefont{{Mamajek}}},
  \bibinfo{journal}{\apj} \textbf{\bibinfo{volume}{758}}, \bibinfo{eid}{31}
  (\bibinfo{year}{2012}), \eprint{1209.5433}.

\bibitem[{\citenamefont{{Preibisch} and {Zinnecker}}(1999)}]{preibisch99}
\bibinfo{author}{\bibfnamefont{T.}~\bibnamefont{{Preibisch}}} \bibnamefont{and}
  \bibinfo{author}{\bibfnamefont{H.}~\bibnamefont{{Zinnecker}}},
  \bibinfo{journal}{\aj} \textbf{\bibinfo{volume}{117}}, \bibinfo{pages}{2381}
  (\bibinfo{year}{1999}).

\bibitem[{\citenamefont{{Ardila} et~al.}(2000)\citenamefont{{Ardila},
  {Mart{\'{\i}}n}, and {Basri}}}]{ardila00}
\bibinfo{author}{\bibfnamefont{D.}~\bibnamefont{{Ardila}}},
  \bibinfo{author}{\bibfnamefont{E.}~\bibnamefont{{Mart{\'{\i}}n}}},
  \bibnamefont{and} \bibinfo{author}{\bibfnamefont{G.}~\bibnamefont{{Basri}}},
  \bibinfo{journal}{\aj} \textbf{\bibinfo{volume}{120}}, \bibinfo{pages}{479}
  (\bibinfo{year}{2000}), \eprint{astro-ph/0003316}.

\bibitem[{\citenamefont{{Slesnick} et~al.}(2006)\citenamefont{{Slesnick},
  {Carpenter}, and {Hillenbrand}}}]{slesnick06usco}
\bibinfo{author}{\bibfnamefont{C.~L.} \bibnamefont{{Slesnick}}},
  \bibinfo{author}{\bibfnamefont{J.~M.} \bibnamefont{{Carpenter}}},
  \bibnamefont{and} \bibinfo{author}{\bibfnamefont{L.~A.}
  \bibnamefont{{Hillenbrand}}}, \bibinfo{journal}{\aj}
  \textbf{\bibinfo{volume}{131}}, \bibinfo{pages}{3016} (\bibinfo{year}{2006}),
  \eprint{astro-ph/0602298}.

\bibitem[{\citenamefont{{Skrutskie} et~al.}(2006)\citenamefont{{Skrutskie},
  {Cutri}, {Stiening}, {Weinberg}, {Schneider}, {Carpenter}, {Beichman},
  {Capps}, {Chester}, {Elias} et~al.}}]{skrutskie06}
\bibinfo{author}{\bibfnamefont{M.~F.} \bibnamefont{{Skrutskie}}},
  \bibinfo{author}{\bibfnamefont{R.~M.} \bibnamefont{{Cutri}}},
  \bibinfo{author}{\bibfnamefont{R.}~\bibnamefont{{Stiening}}},
  \bibinfo{author}{\bibfnamefont{M.~D.} \bibnamefont{{Weinberg}}},
  \bibinfo{author}{\bibfnamefont{S.}~\bibnamefont{{Schneider}}},
  \bibinfo{author}{\bibfnamefont{J.~M.} \bibnamefont{{Carpenter}}},
  \bibinfo{author}{\bibfnamefont{C.}~\bibnamefont{{Beichman}}},
  \bibinfo{author}{\bibfnamefont{R.}~\bibnamefont{{Capps}}},
  \bibinfo{author}{\bibfnamefont{T.}~\bibnamefont{{Chester}}},
  \bibinfo{author}{\bibfnamefont{J.}~\bibnamefont{{Elias}}},
  \bibnamefont{et~al.}, \bibinfo{journal}{\aj} \textbf{\bibinfo{volume}{131}},
  \bibinfo{pages}{1163} (\bibinfo{year}{2006}).

\bibitem[{\citenamefont{{Cutri} and {et al.}}(2013)}]{cutri13}
\bibinfo{author}{\bibfnamefont{R.~M.} \bibnamefont{{Cutri}}} \bibnamefont{and}
  \bibinfo{author}{\bibnamefont{{et al.}}}, \bibinfo{journal}{VizieR Online
  Data Catalog} \textbf{\bibinfo{volume}{2328}} (\bibinfo{year}{2013}).

\bibitem[{\citenamefont{{Henden} et~al.}(2012)\citenamefont{{Henden}, {Levine},
  {Terrell}, {Smith}, and {Welch}}}]{henden12}
\bibinfo{author}{\bibfnamefont{A.~A.} \bibnamefont{{Henden}}},
  \bibinfo{author}{\bibfnamefont{S.~E.} \bibnamefont{{Levine}}},
  \bibinfo{author}{\bibfnamefont{D.}~\bibnamefont{{Terrell}}},
  \bibinfo{author}{\bibfnamefont{T.~C.} \bibnamefont{{Smith}}},
  \bibnamefont{and} \bibinfo{author}{\bibfnamefont{D.}~\bibnamefont{{Welch}}},
  \bibinfo{journal}{Journal of the American Association of Variable Star
  Observers (JAAVSO)} \textbf{\bibinfo{volume}{40}}, \bibinfo{pages}{430}
  (\bibinfo{year}{2012}).

\bibitem[{\citenamefont{{Gaia Collaboration} et~al.}(2016)\citenamefont{{Gaia
  Collaboration}, {Brown}, {Vallenari}, {Prusti}, {de Bruijne}, {Mignard},
  {Drimmel}, and {co-authors}}}]{gaia2016dr}
\bibinfo{author}{\bibnamefont{{Gaia Collaboration}}},
  \bibinfo{author}{\bibfnamefont{A.~G.~A.} \bibnamefont{{Brown}}},
  \bibinfo{author}{\bibfnamefont{A.}~\bibnamefont{{Vallenari}}},
  \bibinfo{author}{\bibfnamefont{T.}~\bibnamefont{{Prusti}}},
  \bibinfo{author}{\bibfnamefont{J.}~\bibnamefont{{de Bruijne}}},
  \bibinfo{author}{\bibfnamefont{F.}~\bibnamefont{{Mignard}}},
  \bibinfo{author}{\bibfnamefont{R.}~\bibnamefont{{Drimmel}}},
  \bibnamefont{and}
  \bibinfo{author}{\bibfnamefont{.}~\bibnamefont{{co-authors}}},
  \bibinfo{journal}{ArXiv e-prints}  (\bibinfo{year}{2016}),
  \eprint{1609.04172}.

\bibitem[{\citenamefont{{Zacharias} et~al.}(2013)\citenamefont{{Zacharias},
  {Finch}, {Girard}, {Henden}, {Bartlett}, {Monet}, and
  {Zacharias}}}]{zacharias13}
\bibinfo{author}{\bibfnamefont{N.}~\bibnamefont{{Zacharias}}},
  \bibinfo{author}{\bibfnamefont{C.~T.} \bibnamefont{{Finch}}},
  \bibinfo{author}{\bibfnamefont{T.~M.} \bibnamefont{{Girard}}},
  \bibinfo{author}{\bibfnamefont{A.}~\bibnamefont{{Henden}}},
  \bibinfo{author}{\bibfnamefont{J.~L.} \bibnamefont{{Bartlett}}},
  \bibinfo{author}{\bibfnamefont{D.~G.} \bibnamefont{{Monet}}},
  \bibnamefont{and} \bibinfo{author}{\bibfnamefont{M.~I.}
  \bibnamefont{{Zacharias}}}, \bibinfo{journal}{\aj}
  \textbf{\bibinfo{volume}{145}}, \bibinfo{eid}{44} (\bibinfo{year}{2013}),
  \eprint{1212.6182}.

\bibitem[{\citenamefont{{Roeser} et~al.}(2010)\citenamefont{{Roeser},
  {Demleitner}, and {Schilbach}}}]{roeser10}
\bibinfo{author}{\bibfnamefont{S.}~\bibnamefont{{Roeser}}},
  \bibinfo{author}{\bibfnamefont{M.}~\bibnamefont{{Demleitner}}},
  \bibnamefont{and}
  \bibinfo{author}{\bibfnamefont{E.}~\bibnamefont{{Schilbach}}},
  \bibinfo{journal}{\aj} \textbf{\bibinfo{volume}{139}}, \bibinfo{pages}{2440}
  (\bibinfo{year}{2010}), \eprint{1003.5852}.

\bibitem[{\citenamefont{{Girard} et~al.}(2011)\citenamefont{{Girard}, {van
  Altena}, {Zacharias}, {Vieira}, {Casetti-Dinescu}, {Castillo}, {Herrera},
  {Lee}, {Beers}, {Monet} et~al.}}]{girard11}
\bibinfo{author}{\bibfnamefont{T.~M.} \bibnamefont{{Girard}}},
  \bibinfo{author}{\bibfnamefont{W.~F.} \bibnamefont{{van Altena}}},
  \bibinfo{author}{\bibfnamefont{N.}~\bibnamefont{{Zacharias}}},
  \bibinfo{author}{\bibfnamefont{K.}~\bibnamefont{{Vieira}}},
  \bibinfo{author}{\bibfnamefont{D.~I.} \bibnamefont{{Casetti-Dinescu}}},
  \bibinfo{author}{\bibfnamefont{D.}~\bibnamefont{{Castillo}}},
  \bibinfo{author}{\bibfnamefont{D.}~\bibnamefont{{Herrera}}},
  \bibinfo{author}{\bibfnamefont{Y.~S.} \bibnamefont{{Lee}}},
  \bibinfo{author}{\bibfnamefont{T.~C.} \bibnamefont{{Beers}}},
  \bibinfo{author}{\bibfnamefont{D.~G.} \bibnamefont{{Monet}}},
  \bibnamefont{et~al.}, \bibinfo{journal}{\aj} \textbf{\bibinfo{volume}{142}},
  \bibinfo{eid}{15} (\bibinfo{year}{2011}), \eprint{1104.5708}.

\bibitem[{\citenamefont{{Arenou} et~al.}(2017)\citenamefont{{Arenou}, {Luri},
  {Babusiaux}, {Fabricius}, {Helmi}, {Robin}, {Vallenari}, {Blanco-Cuaresma},
  {Cantat-Gaudin}, {Findeisen} et~al.}}]{arenou17}
\bibinfo{author}{\bibfnamefont{F.}~\bibnamefont{{Arenou}}},
  \bibinfo{author}{\bibfnamefont{X.}~\bibnamefont{{Luri}}},
  \bibinfo{author}{\bibfnamefont{C.}~\bibnamefont{{Babusiaux}}},
  \bibinfo{author}{\bibfnamefont{C.}~\bibnamefont{{Fabricius}}},
  \bibinfo{author}{\bibfnamefont{A.}~\bibnamefont{{Helmi}}},
  \bibinfo{author}{\bibfnamefont{A.~C.} \bibnamefont{{Robin}}},
  \bibinfo{author}{\bibfnamefont{A.}~\bibnamefont{{Vallenari}}},
  \bibinfo{author}{\bibfnamefont{S.}~\bibnamefont{{Blanco-Cuaresma}}},
  \bibinfo{author}{\bibfnamefont{T.}~\bibnamefont{{Cantat-Gaudin}}},
  \bibinfo{author}{\bibfnamefont{K.}~\bibnamefont{{Findeisen}}},
  \bibnamefont{et~al.}, \bibinfo{journal}{ArXiv e-prints}
  (\bibinfo{year}{2017}), \eprint{1701.00292}.

\bibitem[{\citenamefont{{Pecaut} and {Mamajek}}(2013)}]{pecaut13}
\bibinfo{author}{\bibfnamefont{M.~J.} \bibnamefont{{Pecaut}}} \bibnamefont{and}
  \bibinfo{author}{\bibfnamefont{E.~E.} \bibnamefont{{Mamajek}}},
  \bibinfo{journal}{\apjs} \textbf{\bibinfo{volume}{208}}, \bibinfo{eid}{9}
  (\bibinfo{year}{2013}), \eprint{1307.2657}.

\bibitem[{\citenamefont{{Fiorucci} and {Munari}}(2003)}]{fiorucci03}
\bibinfo{author}{\bibfnamefont{M.}~\bibnamefont{{Fiorucci}}} \bibnamefont{and}
  \bibinfo{author}{\bibfnamefont{U.}~\bibnamefont{{Munari}}},
  \bibinfo{journal}{\aap} \textbf{\bibinfo{volume}{401}}, \bibinfo{pages}{781}
  (\bibinfo{year}{2003}).

\bibitem[{\citenamefont{{Stassun} et~al.}(2014)\citenamefont{{Stassun},
  {Feiden}, and {Torres}}}]{stassun14}
\bibinfo{author}{\bibfnamefont{K.~G.} \bibnamefont{{Stassun}}},
  \bibinfo{author}{\bibfnamefont{G.~A.} \bibnamefont{{Feiden}}},
  \bibnamefont{and} \bibinfo{author}{\bibfnamefont{G.}~\bibnamefont{{Torres}}},
  \bibinfo{journal}{NewAR} \textbf{\bibinfo{volume}{60}}, \bibinfo{pages}{1}
  (\bibinfo{year}{2014}), \eprint{1406.3788}.

\bibitem[{\citenamefont{{Siess} et~al.}(2000)\citenamefont{{Siess}, {Dufour},
  and {Forestini}}}]{siess00}
\bibinfo{author}{\bibfnamefont{L.}~\bibnamefont{{Siess}}},
  \bibinfo{author}{\bibfnamefont{E.}~\bibnamefont{{Dufour}}}, \bibnamefont{and}
  \bibinfo{author}{\bibfnamefont{M.}~\bibnamefont{{Forestini}}},
  \bibinfo{journal}{\aap} \textbf{\bibinfo{volume}{358}}, \bibinfo{pages}{593}
  (\bibinfo{year}{2000}), \eprint{astro-ph/0003477}.

\bibitem[{\citenamefont{{Dotter} et~al.}(2008)\citenamefont{{Dotter},
  {Chaboyer}, {Jevremovi{\'c}}, {Kostov}, {Baron}, and {Ferguson}}}]{dotter08}
\bibinfo{author}{\bibfnamefont{A.}~\bibnamefont{{Dotter}}},
  \bibinfo{author}{\bibfnamefont{B.}~\bibnamefont{{Chaboyer}}},
  \bibinfo{author}{\bibfnamefont{D.}~\bibnamefont{{Jevremovi{\'c}}}},
  \bibinfo{author}{\bibfnamefont{V.}~\bibnamefont{{Kostov}}},
  \bibinfo{author}{\bibfnamefont{E.}~\bibnamefont{{Baron}}}, \bibnamefont{and}
  \bibinfo{author}{\bibfnamefont{J.~W.} \bibnamefont{{Ferguson}}},
  \bibinfo{journal}{\apjs} \textbf{\bibinfo{volume}{178}},
  \bibinfo{eid}{89-101} (\bibinfo{year}{2008}), \eprint{0804.4473}.

\bibitem[{\citenamefont{{Chabrier}}(2003)}]{chabrier03}
\bibinfo{author}{\bibfnamefont{G.}~\bibnamefont{{Chabrier}}},
  \bibinfo{journal}{\pasp} \textbf{\bibinfo{volume}{115}}, \bibinfo{pages}{763}
  (\bibinfo{year}{2003}), \eprint{astro-ph/0304382}.

\bibitem[{\citenamefont{{Jose} et~al.}(2016{\natexlab{a}})\citenamefont{{Jose},
  {Herczeg}, {Samal}, {Fang}, and {Panwar}}}]{jose17}
\bibinfo{author}{\bibfnamefont{J.}~\bibnamefont{{Jose}}},
  \bibinfo{author}{\bibfnamefont{G.~J.} \bibnamefont{{Herczeg}}},
  \bibinfo{author}{\bibfnamefont{M.~R.} \bibnamefont{{Samal}}},
  \bibinfo{author}{\bibfnamefont{Q.}~\bibnamefont{{Fang}}}, \bibnamefont{and}
  \bibinfo{author}{\bibfnamefont{N.}~\bibnamefont{{Panwar}}},
  \bibinfo{journal}{ArXiv e-prints}  (\bibinfo{year}{2016}{\natexlab{a}}),
  \eprint{1612.00697}.

\bibitem[{\citenamefont{{Reggiani} et~al.}(2011)\citenamefont{{Reggiani},
  {Robberto}, {Da Rio}, {Meyer}, {Soderblom}, and {Ricci}}}]{reggiani11}
\bibinfo{author}{\bibfnamefont{M.}~\bibnamefont{{Reggiani}}},
  \bibinfo{author}{\bibfnamefont{M.}~\bibnamefont{{Robberto}}},
  \bibinfo{author}{\bibfnamefont{N.}~\bibnamefont{{Da Rio}}},
  \bibinfo{author}{\bibfnamefont{M.~R.} \bibnamefont{{Meyer}}},
  \bibinfo{author}{\bibfnamefont{D.~R.} \bibnamefont{{Soderblom}}},
  \bibnamefont{and} \bibinfo{author}{\bibfnamefont{L.}~\bibnamefont{{Ricci}}},
  \bibinfo{journal}{\aap} \textbf{\bibinfo{volume}{534}}, \bibinfo{eid}{A83}
  (\bibinfo{year}{2011}), \eprint{1108.1015}.

\bibitem[{\citenamefont{{Herczeg} and {Hillenbrand}}(2014)}]{herczeg14}
\bibinfo{author}{\bibfnamefont{G.~J.} \bibnamefont{{Herczeg}}}
  \bibnamefont{and} \bibinfo{author}{\bibfnamefont{L.~A.}
  \bibnamefont{{Hillenbrand}}}, \bibinfo{journal}{\apj}
  \textbf{\bibinfo{volume}{786}}, \bibinfo{eid}{97} (\bibinfo{year}{2014}),
  \eprint{1403.1675}.

\bibitem[{\citenamefont{{Lu} et~al.}(2013)\citenamefont{{Lu}, {Do}, {Ghez},
  {Morris}, {Yelda}, and {Matthews}}}]{lu13}
\bibinfo{author}{\bibfnamefont{J.~R.} \bibnamefont{{Lu}}},
  \bibinfo{author}{\bibfnamefont{T.}~\bibnamefont{{Do}}},
  \bibinfo{author}{\bibfnamefont{A.~M.} \bibnamefont{{Ghez}}},
  \bibinfo{author}{\bibfnamefont{M.~R.} \bibnamefont{{Morris}}},
  \bibinfo{author}{\bibfnamefont{S.}~\bibnamefont{{Yelda}}}, \bibnamefont{and}
  \bibinfo{author}{\bibfnamefont{K.}~\bibnamefont{{Matthews}}},
  \bibinfo{journal}{\apj} \textbf{\bibinfo{volume}{764}}, \bibinfo{eid}{155}
  (\bibinfo{year}{2013}), \eprint{1301.0540}.

\bibitem[{\citenamefont{{Kouwenhoven} et~al.}(2007)\citenamefont{{Kouwenhoven},
  {Brown}, {Portegies Zwart}, and {Kaper}}}]{kouwenhoven07}
\bibinfo{author}{\bibfnamefont{M.~B.~N.} \bibnamefont{{Kouwenhoven}}},
  \bibinfo{author}{\bibfnamefont{A.~G.~A.} \bibnamefont{{Brown}}},
  \bibinfo{author}{\bibfnamefont{S.~F.} \bibnamefont{{Portegies Zwart}}},
  \bibnamefont{and} \bibinfo{author}{\bibfnamefont{L.}~\bibnamefont{{Kaper}}},
  \bibinfo{journal}{\aap} \textbf{\bibinfo{volume}{474}}, \bibinfo{pages}{77}
  (\bibinfo{year}{2007}), \eprint{0707.2746}.

\bibitem[{\citenamefont{{Kobulnicky} and {Fryer}}(2007)}]{kobulnicky07}
\bibinfo{author}{\bibfnamefont{H.~A.} \bibnamefont{{Kobulnicky}}}
  \bibnamefont{and} \bibinfo{author}{\bibfnamefont{C.~L.}
  \bibnamefont{{Fryer}}}, \bibinfo{journal}{\apj}
  \textbf{\bibinfo{volume}{670}}, \bibinfo{pages}{747} (\bibinfo{year}{2007}).

\bibitem[{\citenamefont{{Lafreni{\`e}re}
  et~al.}(2008)\citenamefont{{Lafreni{\`e}re}, {Jayawardhana}, {Brandeker},
  {Ahmic}, and {van Kerkwijk}}}]{lafreniere08}
\bibinfo{author}{\bibfnamefont{D.}~\bibnamefont{{Lafreni{\`e}re}}},
  \bibinfo{author}{\bibfnamefont{R.}~\bibnamefont{{Jayawardhana}}},
  \bibinfo{author}{\bibfnamefont{A.}~\bibnamefont{{Brandeker}}},
  \bibinfo{author}{\bibfnamefont{M.}~\bibnamefont{{Ahmic}}}, \bibnamefont{and}
  \bibinfo{author}{\bibfnamefont{M.~H.} \bibnamefont{{van Kerkwijk}}},
  \bibinfo{journal}{\apj} \textbf{\bibinfo{volume}{683}},
  \bibinfo{eid}{844-861} (\bibinfo{year}{2008}), \eprint{0803.0561}.

\bibitem[{\citenamefont{{Reggiani} and {Meyer}}(2013)}]{reggiani13}
\bibinfo{author}{\bibfnamefont{M.}~\bibnamefont{{Reggiani}}} \bibnamefont{and}
  \bibinfo{author}{\bibfnamefont{M.~R.} \bibnamefont{{Meyer}}},
  \bibinfo{journal}{\aap} \textbf{\bibinfo{volume}{553}}, \bibinfo{eid}{A124}
  (\bibinfo{year}{2013}), \eprint{1304.3459}.

\bibitem[{\citenamefont{{Kouwenhoven} et~al.}(2009)\citenamefont{{Kouwenhoven},
  {Brown}, {Goodwin}, {Portegies Zwart}, and {Kaper}}}]{kouwenhoven09}
\bibinfo{author}{\bibfnamefont{M.~B.~N.} \bibnamefont{{Kouwenhoven}}},
  \bibinfo{author}{\bibfnamefont{A.~G.~A.} \bibnamefont{{Brown}}},
  \bibinfo{author}{\bibfnamefont{S.~P.} \bibnamefont{{Goodwin}}},
  \bibinfo{author}{\bibfnamefont{S.~F.} \bibnamefont{{Portegies Zwart}}},
  \bibnamefont{and} \bibinfo{author}{\bibfnamefont{L.}~\bibnamefont{{Kaper}}},
  \bibinfo{journal}{\aap} \textbf{\bibinfo{volume}{493}}, \bibinfo{pages}{979}
  (\bibinfo{year}{2009}), \eprint{0811.2859}.

\bibitem[{\citenamefont{{Palla} and {Stahler}}(2000)}]{palla00}
\bibinfo{author}{\bibfnamefont{F.}~\bibnamefont{{Palla}}} \bibnamefont{and}
  \bibinfo{author}{\bibfnamefont{S.~W.} \bibnamefont{{Stahler}}},
  \bibinfo{journal}{\apj} \textbf{\bibinfo{volume}{540}}, \bibinfo{pages}{255}
  (\bibinfo{year}{2000}).

\bibitem[{\citenamefont{{de Geus}}(1992)}]{degeus92}
\bibinfo{author}{\bibfnamefont{E.~J.} \bibnamefont{{de Geus}}},
  \bibinfo{journal}{\aap} \textbf{\bibinfo{volume}{262}}, \bibinfo{pages}{258}
  (\bibinfo{year}{1992}).

\bibitem[{\citenamefont{{Mamajek} et~al.}(2013)\citenamefont{{Mamajek},
  {Pecaut}, {Nguyen}, and {Bubar}}}]{mamajek13}
\bibinfo{author}{\bibfnamefont{E.~E.} \bibnamefont{{Mamajek}}},
  \bibinfo{author}{\bibfnamefont{M.~J.} \bibnamefont{{Pecaut}}},
  \bibinfo{author}{\bibfnamefont{D.~C.} \bibnamefont{{Nguyen}}},
  \bibnamefont{and} \bibinfo{author}{\bibfnamefont{E.~J.}
  \bibnamefont{{Bubar}}}, in \emph{\bibinfo{booktitle}{Protostars and Planets
  VI Posters}} (\bibinfo{year}{2013}).

\bibitem[{\citenamefont{{Ekstr{\"o}m} et~al.}(2012)\citenamefont{{Ekstr{\"o}m},
  {Georgy}, {Eggenberger}, {Meynet}, {Mowlavi}, {Wyttenbach}, {Granada},
  {Decressin}, {Hirschi}, {Frischknecht} et~al.}}]{ekstrom12}
\bibinfo{author}{\bibfnamefont{S.}~\bibnamefont{{Ekstr{\"o}m}}},
  \bibinfo{author}{\bibfnamefont{C.}~\bibnamefont{{Georgy}}},
  \bibinfo{author}{\bibfnamefont{P.}~\bibnamefont{{Eggenberger}}},
  \bibinfo{author}{\bibfnamefont{G.}~\bibnamefont{{Meynet}}},
  \bibinfo{author}{\bibfnamefont{N.}~\bibnamefont{{Mowlavi}}},
  \bibinfo{author}{\bibfnamefont{A.}~\bibnamefont{{Wyttenbach}}},
  \bibinfo{author}{\bibfnamefont{A.}~\bibnamefont{{Granada}}},
  \bibinfo{author}{\bibfnamefont{T.}~\bibnamefont{{Decressin}}},
  \bibinfo{author}{\bibfnamefont{R.}~\bibnamefont{{Hirschi}}},
  \bibinfo{author}{\bibfnamefont{U.}~\bibnamefont{{Frischknecht}}},
  \bibnamefont{et~al.}, \bibinfo{journal}{\aap} \textbf{\bibinfo{volume}{537}},
  \bibinfo{eid}{A146} (\bibinfo{year}{2012}), \eprint{1110.5049}.

\bibitem[{\citenamefont{{Dale} et~al.}(2012)\citenamefont{{Dale}, {Ercolano},
  and {Bonnell}}}]{dale12}
\bibinfo{author}{\bibfnamefont{J.~E.} \bibnamefont{{Dale}}},
  \bibinfo{author}{\bibfnamefont{B.}~\bibnamefont{{Ercolano}}},
  \bibnamefont{and} \bibinfo{author}{\bibfnamefont{I.~A.}
  \bibnamefont{{Bonnell}}}, \bibinfo{journal}{\mnras}
  \textbf{\bibinfo{volume}{424}}, \bibinfo{pages}{377} (\bibinfo{year}{2012}),
  \eprint{1205.0360}.

\bibitem[{\citenamefont{{Bik} et~al.}(2012)\citenamefont{{Bik}, {Henning},
  {Stolte}, {Brandner}, {Gouliermis}, {Gennaro}, {Pasquali}, {Rochau},
  {Beuther}, {Ageorges} et~al.}}]{bik12}
\bibinfo{author}{\bibfnamefont{A.}~\bibnamefont{{Bik}}},
  \bibinfo{author}{\bibfnamefont{T.}~\bibnamefont{{Henning}}},
  \bibinfo{author}{\bibfnamefont{A.}~\bibnamefont{{Stolte}}},
  \bibinfo{author}{\bibfnamefont{W.}~\bibnamefont{{Brandner}}},
  \bibinfo{author}{\bibfnamefont{D.~A.} \bibnamefont{{Gouliermis}}},
  \bibinfo{author}{\bibfnamefont{M.}~\bibnamefont{{Gennaro}}},
  \bibinfo{author}{\bibfnamefont{A.}~\bibnamefont{{Pasquali}}},
  \bibinfo{author}{\bibfnamefont{B.}~\bibnamefont{{Rochau}}},
  \bibinfo{author}{\bibfnamefont{H.}~\bibnamefont{{Beuther}}},
  \bibinfo{author}{\bibfnamefont{N.}~\bibnamefont{{Ageorges}}},
  \bibnamefont{et~al.}, \bibinfo{journal}{\apj} \textbf{\bibinfo{volume}{744}},
  \bibinfo{eid}{87} (\bibinfo{year}{2012}), \eprint{1109.3467}.

\bibitem[{\citenamefont{{Rivera-Ingraham}
  et~al.}(2013)\citenamefont{{Rivera-Ingraham}, {Martin}, {Polychroni},
  {Motte}, {Schneider}, {Bontemps}, {Hennemann}, {Men'shchikov}, {Nguyen
  Luong}, {Andr{\'e}} et~al.}}]{rivera13}
\bibinfo{author}{\bibfnamefont{A.}~\bibnamefont{{Rivera-Ingraham}}},
  \bibinfo{author}{\bibfnamefont{P.~G.} \bibnamefont{{Martin}}},
  \bibinfo{author}{\bibfnamefont{D.}~\bibnamefont{{Polychroni}}},
  \bibinfo{author}{\bibfnamefont{F.}~\bibnamefont{{Motte}}},
  \bibinfo{author}{\bibfnamefont{N.}~\bibnamefont{{Schneider}}},
  \bibinfo{author}{\bibfnamefont{S.}~\bibnamefont{{Bontemps}}},
  \bibinfo{author}{\bibfnamefont{M.}~\bibnamefont{{Hennemann}}},
  \bibinfo{author}{\bibfnamefont{A.}~\bibnamefont{{Men'shchikov}}},
  \bibinfo{author}{\bibfnamefont{Q.}~\bibnamefont{{Nguyen Luong}}},
  \bibinfo{author}{\bibfnamefont{P.}~\bibnamefont{{Andr{\'e}}}},
  \bibnamefont{et~al.}, \bibinfo{journal}{\apj} \textbf{\bibinfo{volume}{766}},
  \bibinfo{eid}{85} (\bibinfo{year}{2013}), \eprint{1301.3805}.

\bibitem[{\citenamefont{{Kiminki} et~al.}(2015)\citenamefont{{Kiminki}, {Kim},
  {Bagley}, {Sherry}, and {Rieke}}}]{kiminki15}
\bibinfo{author}{\bibfnamefont{M.~M.} \bibnamefont{{Kiminki}}},
  \bibinfo{author}{\bibfnamefont{J.~S.} \bibnamefont{{Kim}}},
  \bibinfo{author}{\bibfnamefont{M.~B.} \bibnamefont{{Bagley}}},
  \bibinfo{author}{\bibfnamefont{W.~H.} \bibnamefont{{Sherry}}},
  \bibnamefont{and} \bibinfo{author}{\bibfnamefont{G.~H.}
  \bibnamefont{{Rieke}}}, \bibinfo{journal}{\apj}
  \textbf{\bibinfo{volume}{813}}, \bibinfo{eid}{42} (\bibinfo{year}{2015}),
  \eprint{1509.00081}.

\bibitem[{\citenamefont{{Jose} et~al.}(2016{\natexlab{b}})\citenamefont{{Jose},
  {Kim}, {Herczeg}, {Samal}, {Bieging}, {Meyer}, and {Sherry}}}]{jose16}
\bibinfo{author}{\bibfnamefont{J.}~\bibnamefont{{Jose}}},
  \bibinfo{author}{\bibfnamefont{J.~S.} \bibnamefont{{Kim}}},
  \bibinfo{author}{\bibfnamefont{G.~J.} \bibnamefont{{Herczeg}}},
  \bibinfo{author}{\bibfnamefont{M.~R.} \bibnamefont{{Samal}}},
  \bibinfo{author}{\bibfnamefont{J.~H.} \bibnamefont{{Bieging}}},
  \bibinfo{author}{\bibfnamefont{M.~R.} \bibnamefont{{Meyer}}},
  \bibnamefont{and} \bibinfo{author}{\bibfnamefont{W.~H.}
  \bibnamefont{{Sherry}}}, \bibinfo{journal}{\apj}
  \textbf{\bibinfo{volume}{822}}, \bibinfo{eid}{49}
  (\bibinfo{year}{2016}{\natexlab{b}}), \eprint{1602.06212}.

\bibitem[{\citenamefont{{Rom{\'a}n-Z{\'u}{\~n}iga}
  et~al.}(2015)\citenamefont{{Rom{\'a}n-Z{\'u}{\~n}iga}, {Ybarra},
  {Meg{\'{\i}}as}, {Tapia}, {Lada}, and {Alves}}}]{roman15}
\bibinfo{author}{\bibfnamefont{C.~G.}
  \bibnamefont{{Rom{\'a}n-Z{\'u}{\~n}iga}}},
  \bibinfo{author}{\bibfnamefont{J.~E.} \bibnamefont{{Ybarra}}},
  \bibinfo{author}{\bibfnamefont{G.~D.} \bibnamefont{{Meg{\'{\i}}as}}},
  \bibinfo{author}{\bibfnamefont{M.}~\bibnamefont{{Tapia}}},
  \bibinfo{author}{\bibfnamefont{E.~A.} \bibnamefont{{Lada}}},
  \bibnamefont{and} \bibinfo{author}{\bibfnamefont{J.~F.}
  \bibnamefont{{Alves}}}, \bibinfo{journal}{\aj}
  \textbf{\bibinfo{volume}{150}}, \bibinfo{eid}{80} (\bibinfo{year}{2015}),
  \eprint{1507.00016}.

\bibitem[{\citenamefont{{Wu} et~al.}(2016)\citenamefont{{Wu}, {Bik},
  {Bestenlehner}, {Henning}, {Pasquali}, {Brandner}, and {Stolte}}}]{wu16}
\bibinfo{author}{\bibfnamefont{S.-W.} \bibnamefont{{Wu}}},
  \bibinfo{author}{\bibfnamefont{A.}~\bibnamefont{{Bik}}},
  \bibinfo{author}{\bibfnamefont{J.~M.} \bibnamefont{{Bestenlehner}}},
  \bibinfo{author}{\bibfnamefont{T.}~\bibnamefont{{Henning}}},
  \bibinfo{author}{\bibfnamefont{A.}~\bibnamefont{{Pasquali}}},
  \bibinfo{author}{\bibfnamefont{W.}~\bibnamefont{{Brandner}}},
  \bibnamefont{and} \bibinfo{author}{\bibfnamefont{A.}~\bibnamefont{{Stolte}}},
  \bibinfo{journal}{\aap} \textbf{\bibinfo{volume}{589}}, \bibinfo{eid}{A16}
  (\bibinfo{year}{2016}), \eprint{1602.05190}.

\bibitem[{\citenamefont{{Gully-Santiago}
  et~al.}(2017)\citenamefont{{Gully-Santiago}, {Herczeg}, {Czekala}, {Somers},
  {Grankin}, {Covey}, {Donati}, {Alencar}, {Hussain}, {Shappee}
  et~al.}}]{gully17}
\bibinfo{author}{\bibfnamefont{M.~A.} \bibnamefont{{Gully-Santiago}}},
  \bibinfo{author}{\bibfnamefont{G.~J.} \bibnamefont{{Herczeg}}},
  \bibinfo{author}{\bibfnamefont{I.}~\bibnamefont{{Czekala}}},
  \bibinfo{author}{\bibfnamefont{G.}~\bibnamefont{{Somers}}},
  \bibinfo{author}{\bibfnamefont{K.}~\bibnamefont{{Grankin}}},
  \bibinfo{author}{\bibfnamefont{K.~R.} \bibnamefont{{Covey}}},
  \bibinfo{author}{\bibfnamefont{J.~F.} \bibnamefont{{Donati}}},
  \bibinfo{author}{\bibfnamefont{S.~H.~P.} \bibnamefont{{Alencar}}},
  \bibinfo{author}{\bibfnamefont{G.~A.~J.} \bibnamefont{{Hussain}}},
  \bibinfo{author}{\bibfnamefont{B.~J.} \bibnamefont{{Shappee}}},
  \bibnamefont{et~al.}, \bibinfo{journal}{\apj} \textbf{\bibinfo{volume}{836}},
  \bibinfo{eid}{200} (\bibinfo{year}{2017}), \eprint{1701.06703}.

\bibitem[{\citenamefont{{Fang} et~al.}(2016)\citenamefont{{Fang}, {Zhao},
  {Zhao}, {Chen}, and {Bharat Kumar}}}]{fang16}
\bibinfo{author}{\bibfnamefont{X.-S.} \bibnamefont{{Fang}}},
  \bibinfo{author}{\bibfnamefont{G.}~\bibnamefont{{Zhao}}},
  \bibinfo{author}{\bibfnamefont{J.-K.} \bibnamefont{{Zhao}}},
  \bibinfo{author}{\bibfnamefont{Y.-Q.} \bibnamefont{{Chen}}},
  \bibnamefont{and} \bibinfo{author}{\bibfnamefont{Y.}~\bibnamefont{{Bharat
  Kumar}}}, \bibinfo{journal}{\mnras} \textbf{\bibinfo{volume}{463}},
  \bibinfo{pages}{2494} (\bibinfo{year}{2016}), \eprint{1608.05452}.

\bibitem[{\citenamefont{{Somers} and {Stassun}}(2016)}]{somers16}
\bibinfo{author}{\bibfnamefont{G.}~\bibnamefont{{Somers}}} \bibnamefont{and}
  \bibinfo{author}{\bibfnamefont{K.~G.} \bibnamefont{{Stassun}}},
  \bibinfo{journal}{ArXiv e-prints}  (\bibinfo{year}{2016}),
  \eprint{1609.04841}.

\bibitem[{\citenamefont{{Binks} and {Jeffries}}(2014)}]{binks14}
\bibinfo{author}{\bibfnamefont{A.~S.} \bibnamefont{{Binks}}} \bibnamefont{and}
  \bibinfo{author}{\bibfnamefont{R.~D.} \bibnamefont{{Jeffries}}},
  \bibinfo{journal}{\mnras} \textbf{\bibinfo{volume}{438}},
  \bibinfo{pages}{L11} (\bibinfo{year}{2014}), \eprint{1310.2613}.

\bibitem[{\citenamefont{{Kraus} et~al.}(2014)\citenamefont{{Kraus}, {Shkolnik},
  {Allers}, and {Liu}}}]{kraus14}
\bibinfo{author}{\bibfnamefont{A.~L.} \bibnamefont{{Kraus}}},
  \bibinfo{author}{\bibfnamefont{E.~L.} \bibnamefont{{Shkolnik}}},
  \bibinfo{author}{\bibfnamefont{K.~N.} \bibnamefont{{Allers}}},
  \bibnamefont{and} \bibinfo{author}{\bibfnamefont{M.~C.} \bibnamefont{{Liu}}},
  \bibinfo{journal}{\aj} \textbf{\bibinfo{volume}{147}}, \bibinfo{eid}{146}
  (\bibinfo{year}{2014}), \eprint{1403.0050}.

\end{thebibliography}

\end{CJK*}

\end{document}